\documentclass{ws-ijmpd}
\usepackage[super,compress]{cite}
\usepackage{subfigure}
\usepackage{multirow}
\begin{document}

\markboth{Bagchi}{Pulsar Luminosities}

%
\catchline{}{}{}{}{}
%

\title{LUMINOSITIES OF RADIO PULSARS }

\author{MANJARI BAGCHI}

\address{Department of Physics, West Virginia University, \\
Morgantown, West Virginia 26506
US \\ manjari.bagchi@gmail.com}

\maketitle

\begin{history}
\received{Day Month Year}
\revised{Day Month Year}
\end{history}

\begin{abstract}
Luminosity is an intrinsic property of radio pulsars related to the properties of the magnetospheric plasma and the beam geometry, and inversely proportional to the observing frequency. In traditional models, luminosity has been considered as a function of the spin parameters of pulsars. On the other hand, parameter independent models like power law and lognormal have been also used to fit the observed luminosities. Some of the older studies on pulsar luminosities neglected observational biases, but all of the recent studies tried to model observational effects as accurately as possible. Luminosities of pulsars in globular clusters and in the Galactic disk have been studied separately. Older studies concluded that these two categories of pulsars have different luminosity distributions, but the most recent study concluded that those are the same. This article reviews all significant works on pulsar luminosities and discusses open questions.
\end{abstract}


\ccode{PACS number(s): 97.60.Gb, 97.60.Jd, 97.10.Ri,  97.10.Yp,  98.20.Gm}


\section{Introduction}

Pulsars are rotating magnetized neutron stars which emit beamed electromagnetic radiations from magnetic poles. These objects are amazing tools to test different theories of fundamental physics, including general relativity, alternative theories of gravity, particle physics, nuclear physics, magnetism, plasma physics, etc. Although the emission from these objects ranges from gamma-ray to radio, here I confine myself only to radio pulsars. The first radio pulsar was discovered by Jocelyn Bell in 1967\cite{jocelyn}, and since then, there are new discoveries almost every year. A total of 2213 pulsars are listed in the ATNF pulsar catalogue\cite{atnfcat} (http://www.atnf.csiro.au/research/pulsar/psrcat/expert.html; March 2013 version), out of which 6 are in the small Magellanic cloud, 15 are in the large Magellanic cloud and 144 are in 28 Galactic globular clusters. Spin periods of these pulsars range from 1.4 ms to 12 s. It is commonly believed that pulsars are born with spin periods of around 1 s and then they slow down by loosing their rotational kinetic energy in the form of electromagnetic energy. Fast pulsars are believed to gain rotational kinetic energy as a result of mass transfer from their binary companion\cite{alpar, radsrini} (or past companions, in the cases of isolated fast pulsars). Such pulsars are known as ``recycled pulsars" or ``millisecond pulsars" (MSPs), and slow pulsars are called ``normal" pulsars. Since the exact definition of these categories varies, here I stick to the definition of MSPs as pulsars with spin periods less than 20 ms. Although, pulsars having somewhat higher spin periods (upto around 100 ms) could also gain rotational kinetic energy by accreting matter from their companions; in those cases, I prefer the term ``recycled" pulsars. In this review, I try to summarize what has been learnt so far about radio luminosities of pulsars, which is a very important intrinsic property of pulsars and is directly related to the pulsar emission mechanism, the physics of the magnetosphere and the structure of the beam.

\section{Preliminaries on flux densities and luminosities of radio pulsars}
\label{lab:prelim}
Flux density is the amount of energy received from an astronomical object per unit area (of the observing device) per unit time per unit frequency range. The unit of flux density is watts per square metre per hertz. Radio astronomers use the unit jansky (Jy), where 1 jansky $= 10^{-26}$ watts per square metre per hertz. As pulsars are comparatively faint objects, their flux densities are usually expressed in milliJansky (mJy). Flux density is a measurable quantity, but it is not an intrinsic property of the object. The intrinsic property of the object is the luminosity, which is the amount of energy radiated by the object per unit time. For an object radiating in a spherically symmetric manner, the luminosity can be written as $\mathcal{L} = 4 \pi d^2 \, \int_{\nu_{\rm low}}^{\nu_{\rm up}} S_{\nu} d \nu$, where $d$ is the distance of the object from the observer, $S_{\nu}$ is the flux density at any observing frequency $\nu$, $\nu_{\rm low}$ and $\nu_{\rm up}$ being the lower and upper limits of frequencies over which the object has been observed. But in the case of radio pulsars, the emission is not spherically symmetric; rather, it is beamed from two magnetic poles. So one needs to consider the beaming geometry while defining the luminosity of a radio pulsar as:\cite{lk05}
\begin{equation}
\mathcal{L} = \frac{4 \pi d^2}{\delta} \, {\rm sin^2 }\left(\frac{\rho}{2} \right) \int_{\nu_{\rm low}}^{\nu_{\rm up}} S_{\nu} d \nu.
\label{eq:truelum}
\end{equation} Here $\rho$ is the radius of the emission cone (assumed circular), $\delta = W_{\rm eq}/P_{s}$ is the pulse duty cycle, $P_{s}$ is the spin period of the pulsar, $W_{\rm eq}$ is the equivalent width of the pulse (i.e. the width of a top-hat shaped pulse having the same area and peak flux density as the true profile). As it is usually difficult to determine the values of $\rho$ and $\delta$ reliably\footnote{Moreover, there are different models for the beam structure, like ``nested cone", ``patchy beam" etc., see Lorimer and Kramer\cite{lk05} chapter 3.4.3 and references therein for details. In those cases, Eq. (\ref{eq:truelum}) is not valid.}, the parameter ``pseudoluminosity" is defined as
\begin{equation}\label{eq:pseudolum}
L_{\nu} = S_{\nu}~d^2 \, .
\end{equation} Note that, although luminosity has the dimension of power, i.e. watts, pseudoluminosity has the dimension of watts per hertz. In pulsar astronomy, pseudoluminosity is expressed in units of ${\rm mJy~ kpc^2}$, so physical or true luminosity can be expressed in units of ${\rm mJy~ kpc^2~MHz}$. The common practice is to study pseudoluminosity and refer it as luminosity. I follow this convention in the present article (unless otherwise mentioned explicitly), with the subscript implying the value of the observing frequency in MHz, e.g., $L_{400}$ stands for the pseudoluminosity of a pulsar observed at 400 MHz and $L_{1400}$ stands for the pseudoluminosity of a pulsar observed at 1400 MHz.

\begin{figure}[h]
\centerline{\psfig{file=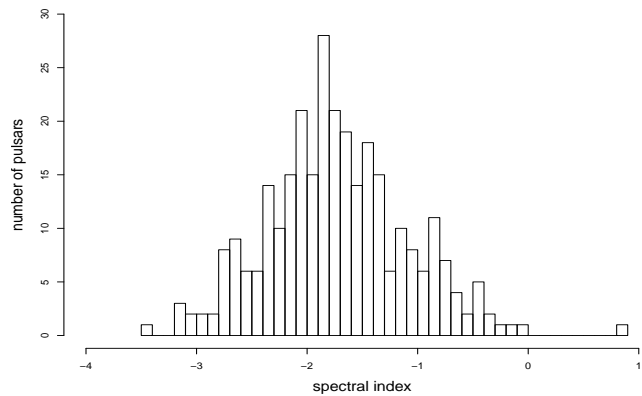,width=6.0cm,height=9.0cm,angle=-90}}
\vspace*{8pt}
\caption{Distribution of spectral indices of 294 pulsars. \label{fig:spectralindex}}
\end{figure}

It is a well known fact that pulsars are not equally bright at different frequencies. $L_{\nu}$ (and as a result $S_{\nu}$) varies with frequency as $L_{\nu} \propto \nu^{\alpha}$ (or $S_{\nu} \propto \nu^{\alpha}$) and $\alpha$ is known as the ``spectral index". The value of $\alpha$ is different for different pulsars. For most of them, it lies in the range of $-1$ to $-2$ implying that pulsars are in general brighter at lower frequencies. In Fig. \ref{fig:spectralindex}, I show the distribution of spectral indices of 294 pulsars listed in the ATNF pulsar catalog (as of March 2013). This distribution has a mean of $-1.68$ and median $-1.70$. Extreme examples are PSR J0711$+$0931 ($\alpha=-3.5$) and PSR J1740$+$1000 ($\alpha=0.9$). Earlier, Maron \textit{et~al.} \cite{mkkw00} found the mean $\alpha$ as $-1.8$ using 281 pulsars. Toscano \textit{et~al.} \cite{tbms98} obtained a mean $\alpha$ of $-1.9$ for 19 millisecond pulsars (out of which two are in globular clusters). Bagchi \textit{et~al.}\cite{manjari11} found a mean $\alpha$ of $-1.9$ for 20 MSPs in globular clusters. They used this value to convert flux densities measured at different frequencies to those at 1400 MHz. They also showed that the luminosity distribution for recycled pulsars in GCs did not vary significantly when $\alpha$ changed in the range of $-1.6$ to $-2.0$. This result agreed with the earlier conclusion by Hessels \textit{et~al.} \cite{hrskf07} using 37 isolated pulsars in globular clusters.

Recently, using Monte-Carlo simulations and taking care of observational limits, Bates \textit{et~al.} \cite{blv13} have found that the true distribution of spectral indices for potentially observable radio pulsars can be modelled as a Gaussian with mean -1.4 and standard deviation 1.

\section{Luminosity law for radio pulsars: function of the spin period and the rate of change of the spin period}
\label{sec:lumlawppdot}

The rate of loss of the rotational kinetic energy of a pulsar can be written as
\begin{equation}
\dot{E} = - \frac{d}{dt} \left( \frac{1}{2} I \Omega^2 \right) = 4 \pi^2 I \, P_{s}^{-3} \, \dot{P}_{s} ~,
\label{eq:spindownlum}
\end{equation} where $\dot{P}_{s}$ is the rate of change of the spin period, $\Omega = 2 \pi / P_{s}$ is the angular frequency and $I$ is the moment of inertia of the pulsar. $\dot{E}$ is called the ``spin-down luminosity" and it is the total power output by the pulsar. In the simplified assumption that pulsars are magnetic dipoles, rotating in vacuum, this power can be equated to the total electromagnetic power emitted by the pulsar (neglecting energy losses in other forms, e.g., gravitational energy, heat energy, etc.). Even in that case, the total power would be distributed over a wide range of frequency. So it is clear that this spin-down luminosity is a quantity different from the radio luminosity we observe, although according to some theoretical models, this $\dot{E}$ is related to the physical radio luminosity $\mathcal{L}$ (Eq \ref{eq:truelum}) or even with the pseudoluminosity $L_{\nu}$.

The luminosity of a radio pulsar at a particular observing frequency $\nu$ is commonly expressed by the functional form:

\begin{equation}
L_{\nu} = a \, P_{s}^{p} \, \dot{P}_{s}^{q}
\label{eq:LLppdot1}
\end{equation} where $a$ is a proportionality constant whose dimension depends upon the values of the indices $p$ and $q$ and the units of $P_{s}$ and $\dot{P}_{s}$. Usually $P_{s}$ is expressed in s and $\dot{P}_{s}$ is ${\rm s \, s^{-1}}$, but sometimes $\dot{P}_{s}$ is given in units of $10^{-15} \, {\rm s \, s^{-1}}$, which changes the value of the constant $a$. Note that when one tries to fit Eq. (\ref{eq:LLppdot1}) to the observed set of pulsars, selection effects play a significant role and the fitted form does not represent the actual population. 

This form was first used by Gunn and Ostriker\cite{go70} who obtained $p=0.905$, $q=0.905$ for the observed $L_{400}$ values of 41 pulsars. This fit can be approximated as $p=1$, $q=1$. They also proposed an exponential decay of the surface magnetic field\footnote{It is now established that the surface magnetic field does not decay\cite{dipankar}.} ($B_S$) with time resulting in a decrease in $L_{400}$ with time as $B_S = 3.2 \times 10^{19} P_{s}^{1/2} \dot{P}_{s}^{1/2}$ ($P_{s}$ in s and $\dot{P}_{s}$ in ${\rm s \, s^{-1}}$) giving $L_{400} \propto B_{S}^{2}$. Lyne, Manchester, and Taylor \cite{lmt85} favoured $L_{400} \propto B_{S}^{2}$ law over few other laws they explored, e.g., $L_{400} \propto \dot{P}_{s}$ and $L_{400} \propto P_{s}^{-1} \dot{P}_{s}$. On the other hand, Lyne, Ritchings, and Smith\cite{lrs75} fitted $L_{400}$ with $p=-1.8$, $q=0.88$ for a sample of 84 pulsars. The same set of pulsars resulted $p=-0.79 \pm 0.30$, $q=0.36 \pm 0.11$ in the study by Vivekanand and Narayan\cite{vn81}. Using a larger set of 242 pulsars, they obtained $p=-0.86 \pm 0.20$, $q=0.38 \pm 0.08$. Later Pr\'oszy\'nski and Przybycie\'n\cite{pp84} found that $p=-1.04 \pm 0.15$, $q= 0.35 \pm 0.06$ using a sample of 275 pulsars. 

Stollman\cite{stoll86} tried to understand the discrepancy between the luminosity law obtained by different people. He found that $p=1$, $q=1$ (chosen by Gunn and Ostriker) explained the distribution of pulsars in the $B_{S}-P$ plane better, but $p=-1$, $q=0.35$ (chosen by Pr\'oszy\'nski and Przybycie\'n) fitted the observed values of $L_{400}$ better. For the first case, he compared a simulated, flux-limited sample of pulsars with the real ones. In his simulation, he chose a normal distribution for the space velocities of pulsars (with a standard deviation of 100 $\rm {km~ s^{-1}}$), a lognormal distribution for initial magnetic fields (with a mean of 12.5 and standard deviation of 0.5), a flat distribution between 1-50 ms for initial spin periods, and an exponential decay law for magnetic fields. Soon he proposed a new luminosity law as\cite{stoll87}:

\begin{subequations}
\begin{align}
  \label{stollmana}       L_{400} & = 10^{-10.05 \pm 0.84} \, \left( B_{S}/P_{s}^2 \right)^{0.98 \pm 0.03} {\rm ~~~ if~} B_{S}/P_{s}^2 \leq 10^{13} ~{\rm G \, s^{-2} } , \\
  \label{stollmanb}      L_{400} & = 10^{2.71 \pm 0.60}{\rm ~~~~~~~~~~~~~~~~~~~~~~~~~~~~~~ if~} B_{S}/P_{s}^2 > 10^{13} ~{\rm G \, s^{-2}}.
\end{align}
\label{stollmanall}
\end{subequations}

Eq (\ref{stollmana}) can be approximated to Eq (\ref{eq:LLppdot1}) with $p=-1.47,~ q=0.49$ which can be further simplified as $p=-1.5,~ q=0.5$. Stollman\cite{stoll87} also gave a physical explanation for this law within the framework of Ruderman-Sutherland\cite{rudsuth75} model assuming $L_{400}$ to be proportional to the potential difference across the polar magnetospheric gap. All the works mentioned so far in this section fitted observed values $L_{400}$ with observed values of $P_{s}$ and $\dot{P}_{s}$, without considering the sensitivity of pulsar surveys properly, and these fits were used to explain/predict other observable parameters. 

Emmering and Chevalier\cite{ec89} fitted $L_{400}$ of observed pulsars by Eq (\ref{eq:LLppdot1}) with $p=-0.96 \pm 0.15$, $q=0.39 \pm 0.06$ which was very close to the fit obtained by Pr\'oszy\'nski and Przybycie\'n. They took care of selection effects and performed Mone-Carlo simulations to get the luminosity function for potentially detectable pulsars (through pulsar surveys performed upto that time), as well as the intrinsic pulsar luminosity function. They assumed that the intrinsic luminosity function had a scatter which could be modelled with a normal distribution with the mean fitted with Eq (\ref{eq:LLppdot1}), and the measured luminosity was the product of this intrinsic distribution and a lognormal distribution. They obtained different values for $p$ and $q$ depending upon the model, e.g., their model I gave $p=-1.11 \pm 0.15$ and $q= 0.43 \pm 0.06$ for detectable pulsars but $p=-1.61$, $q=0.64$ for the intrinsic distribution. Similarly their model II gave $p=-0.82 \pm 0.13$ and $q= 0.31 \pm 0.05$ for detectable pulsars but $p=-1.40$, $q=0.43$ for the intrinsic distribution. On the other hand, Narayan\cite{narayan87}, who also took care of selection effects, fitted the mean luminosity ($L_{400}^{mean}$) with $p=-1$, $q=1/3$, $a=10^{1.72}$ ($\dot{P}_{s}$ was expressed in $10^{-15} \, {\rm s \, s^{-1}}$) with a scatter of the form $0.2144(1+ \cos(1.347X) )$ where $X = \log({L_{400}/L_{400}^{mean}})$. Narayan and Ostriker\cite{narost90} explored different luminosity laws and found that the fit of $L_{400}^{mean}$ for normal, isolated pulsars with $p=1,~q=1/3,~a=10^{1.635}$ ($\dot{P}_{s}$ in $10^{-15} \, {\rm s \, s^{-1}}$) with a scatter of the form $0.0358 \, (X + 1.8)^2 \exp(-3.6X)$ gave the best fit. Kulkarni, Narayan and Romani\cite{nnr90} fitted $L_{400}$ values of recycled pulsars (including both disk and globular cluster pulsars) with $p=1,~q=1/3,~a=10^{1.635}$ (again $\dot{P}_{s}$ was expressed in $10^{-15} \, {\rm s \, s^{-1}}$) with the same scatter, i.e. $0.0358 \, (X + 1.8)^2 \exp(-3.6X)$. For the sake of completeness, I quote two other models used by Narayan and Ostriker\cite{narost90}. The first one was

\begin{equation}
\label{narayanostriker2}
 L_{400}^{mean}   = \left\{ \begin{array}{rl}
 10^{1.613} \, 10^{\frac{1}{2} \log(\dot{P}_{s}/P_{s}^3)}  &\mbox{~if $\dot{P}_{s}/P_{s}^3 < 10^{1.488}$} \\
 10^{2.358}  & \mbox{~if $\dot{P}_{s}/P_{s}^3 > 10^{1.488}$ }  \\
        \end{array} \right.
\end{equation} with a scatter as $0.0335 \, (X + 2.0)^2 \exp(-3.0X)$. The other one was

\begin{equation}
\label{narayanostriker3}
 L_{400}^{mean}   = \left\{ \begin{array}{rl}
 10^{1.458} \, 10^{0.132 \log(\dot{P}_{s}/P_{s}^3)}  &\mbox{~if $\dot{P}_{s}/P_{s}^3 < 10^{0.204}$} \\
 10^{1.317} \, 10^{0.823 \log(\dot{P}_{s}/P_{s}^3)}  & \mbox{~if $10^{1.314}  > \dot{P}_{s}/P_{s}^3 > 10^{0.204}$} \\
  10^{2.398}  & \mbox{~if $\dot{P}_{s}/P_{s}^3 > 10^{1.314}$ } 
       \end{array} \right.
\end{equation} with a scatter as $0.0340 \, (X + 1.9)^2 \exp(-3.3X)$.

Afterwards, with a larger dataset (412 pulsars) and an improved model to estimate the distances of the pulsars, Lorimer \textit{et~al.}\cite{lbdh93} found that neither $p=1,~ q =1$ nor $p=-1,~q=1/3$ adequately described the observed values of $L_{400}$, their best fit values were $p=-0.68 \pm 0.12,~q=0.28 \pm 0.05$. On the other hand, when they used the luminosity laws by Stollman\cite{stoll87} and by Emmering and Chevalier\cite{ec89} with a Gaussian spread of 0.8 as the intrinsic distribution of in $\log L_{400}$, and performed Monte-Carlo simulations, they obtained plausible fits to the $P_{s}$, $B_{S}$, $L_{400}$, $d$ and $DM$ distributions of the observed pulsar populations. Here $DM$ is the dispersion measure, i.e. the free electron column density integrated along the line-of-sight, and is expressed in the unit of ${\rm pc~cm^{-3}}$

Malov, Malov and Malofeev\cite{mmm96} obtained a fit (without considering any selection effect) as
\begin{equation}
 \log L_{400} = (0.64 \pm 0.05) \log (B_{S}/P_{s}^2) + 20.54 \pm 0.68 ~.
 \label{eq:mmm97a}
\end{equation} They also obtained a relation of $L_{400}$ with $B_{LC}/P^2$ where $B_{LC}$ is the magnetic field near the light cylinder, as
\begin{equation}
 \log L_{400} = (1.2 \pm 0.3) \log (B_{LC}/P_{s}^2) + 15.0 \pm 3.5 ~.
 \label{eq:mmm97a}
\end{equation}

Some of the works mentioned in this section, e.g., Narayan\cite{narayan87}, Emmering and Chevalier\cite{ec89}, Narayan and Ostriker\cite{narost90}, Lorimer \textit{et~al.}\cite{lbdh93}, are early examples of population synthesis which has become very popular in pulsar astronomy nowadays. In the case of a population synthesis, first one needs to use appropriate birth distributions of $P_{s}$, $\dot{P}_{s}$ (or $P_{s}$, $B_{S}$), which result a distribution of $L_{\nu}$. Then one needs to model other parameters like the space distribution of pulsars at birth, their motion through the Galactic potential, evolution of the spin parameters, etc. Finally through modeling of pulsar surveys (i.e. detection sensitivity, sky coverage etc.), one obtains synthetic distributions of parameters like $P_{s}$, $\dot{P}_{s}$, $L_{\nu}$ (or $S_{\nu}$) for ``observable" pulsars which are compared with the observed values to test the validity of the models. 

So far, I have discussed only models for pseudoluminosity, but there are efforts to model the physical or true luminosity of radio pulsars as defined in Eq. (\ref{eq:truelum}). Arzoumanian, Chernoff, and Cordes\cite{acc02} modelled 
\begin{equation}
\mathcal{L} = {\rm min} \left\{ \mathcal{L}_{0} P_{s}^{a} \dot{P}_{s}^{b} , \dot{E} \right\}~{\rm ergs \, s^{-1}}
\label{eq:truelumModacc02}
\end{equation} with $\mathcal{L}_{0} = 10^{29.3}$, $a=-1.3$, $b=0.4$ and $\dot{P}_{s}$ in $10^{-15}~{\rm s \, s^{-1}}$. They also assumed that $\mathcal{L} = \epsilon \dot{E}$ where the efficiency factor $\epsilon$ lied in the range of $0.02 - 0.30$. To express $\mathcal{L}$ in the units of ${\rm mJy~kpc^2~MHz}$, one needed to set $\mathcal{L}_{0} = 2.1 \times 10^{12}$. Story \textit{et~al.}\cite{sgh07} studied the emission mechanism and beam geometry in further details and their preferred set of parameters were $\mathcal{L}_{0} = 1.76 \times 10^{10}$, $a=-1.05$, $b=0.37$ and $\dot{P}_{s}$ in $10^{-15}~{\rm s \, s^{-1}}$ for $\mathcal{L}$ in the units of ${\rm mJy~kpc^2~MHz}$.

In the next section, I discuss efforts to understand the distribution function of pseudoluminosities (which I again mention as luminosity) of radio pulsars, independent of spin parameters.

\section{Luminosity function for radio pulsars}

The luminosity function of any specific type of astronomical objects is defined as the number of such objects per unit luminosity interval. This can be done either by using a population synthesis method or by directly fitting the observed luminosities. In this section, first I discuss the direct method and then the population synthesis method used by various people to obtain the luminosity function for pulsars in the Galactic disk. Efforts to obtain the luminosity function for the pulsars in globular clusters will be discussed separately.

\subsection{Direct method}
\label{sublab:diskdirect}

The most popular luminosity function used for radio pulsars is the power law, which can be written as

\begin{equation}
\rho(L_{\nu}) = \rho_0 L_{\nu}^{\gamma}~,
\label{eq:powerlaw1}
\end{equation} where $\rho(L_{\nu})$ is the number or density of pulsars with luminosities in the range of $L_{\nu}$ to $L_{\nu} + dL_{\nu}$ and $\rho_0$ is the constant of proportionality. Commonly, the interval $dL_{\nu}$ is taken as the unit logarithmic interval (as the observed values of $L_{\nu}$ spread over several orders of magnitudes). Large\cite{large71} found that $\gamma = -2$ fitted $L_{408}$ of 29 Molonglo pulsars moderately. But when he considered errors as a result of Poisson distribution, he found that $\gamma$ changed from $-1.5$ to $-3$ with the increase of $L_{408}$. This change in the value of $\gamma$ became even larger when he tried to model the beaming fraction. Roberts\cite{rob76} fitted a truncated power law of $\gamma = -1.7 \pm 0.3$ to the observed $L_{400}$ of the 44 pulsars out of 50 discovered by the pulsar survey performed by Hulse and Taylor\cite{ht75} using the Arecibo radio telescope. Davies, Lyne, and Seiradakis \cite{dls77} fitted $L_{408}$ of 51 pulsars (20 new) detected in the Jodrell Bank pulsar survey with $\gamma = -1.96$. Taylor and Manchester\cite{tm77} fitted observed $L_{400}$ of the 110 pulsars (discovered by the three largest pulsar surveys of that time, the Molonglo survey\cite{lv71}, the Jodrell bank survey\cite{dls77}, and the Arecibo survey\cite{ht75}) with a power law of $\gamma = -2.12 \pm 0.03$ truncated at $L_{400,{\rm min}} = 3~{\rm mJy~kpc^2}$. They also argued that Roberts\cite{rob76} obtained a flatter distribution due to overestimation of the volume sampled at low luminosities. Lyne, Manchester, and Taylor\cite{lmt85} chose $\gamma=0$. Note that the values of individual luminosities and the constant $\rho_0$ in these early works are not of much significance today, because at that time, the distances of the pulsars were not estimated correctly. As an extreme example, Davies, Lyne, and Seiradakis \cite{dls77} defined $L_{400} = S_{400} \, DM^2$. Moreover, in these studies, $\rho(L_{\nu})$ was chosen as the space density of pulsars in the luminosity interval $L_{\nu}$ to $L_{\nu} + dL_{\nu}$, so to obtain the number of pulsars in that interval, integration over the volume was required. But the use of $\rho(L_{\nu})$ as the number of pulsars in the luminosity interval of $L_{\nu}$ to $L_{\nu} + dL_{\nu}$ is equally valid. $\rho(L_{\nu})$ can be converted to a probability distribution function (PDF) with proper normalization, i.e. by adjusting the constant $\rho_0$ and can be written as:
\begin{equation} 
f_{\rm pl} \, (L_{\nu})  = - (\gamma + 1) L_{\rm \nu, min}^{-(\gamma+1)} L_{\nu}^{\gamma} \, ,
\label{eq:singlepow_dist_def}
\end{equation}
where $L_{\rm \nu, min}$ is the minimum value of $L_{\nu}$. Equation (\ref{eq:singlepow_dist_def}) leads to the cumulative distribution function (CDF) as 
\begin{equation} 
F_{\rm pl} \, (L_{\nu}) = \int_{L_{\rm \nu, min}}^{L_{\nu}} \, f_{\rm pl} \, (x) dx = 1 - \left(\frac{L_{\rm \nu, min}}{L_{ \nu}} \right)^{-(\gamma + 1)}  ,
\label{eq:singlepow_cdf}
\end{equation} and the complementary cumulative distribution
function (CCDF) as
\begin{equation} 
F_{ c \, {\rm pl}} \, (L_{\nu})  = 1 - F_{\rm pl} \, (L_{\nu}) = \left(\frac{L_{\rm \nu, min}}{L_{ \nu}} \right)^{-(\gamma + 1)} \, = \left(\frac{L_{\rm \nu}}{L_{\rm \nu, min}} \right)^{(\gamma + 1)}  .
\label{eq:singlepow_ccdf} 
\end{equation}

$F_{ c \, {\rm pl}} \, (L_{\nu})$ is the probability that the value of the radio luminosity at the observing frequency $\nu$ is greater than $L_{\nu}$. But the number of pulsars having luminosities greater than or equal to $L_{\nu}$ should be expressed with the complementary cumulative frequency distribution function (CCFDF) as: 
\begin{equation} 
N(\geq L_{\nu})  = N_{\rm tot} \, \left(\frac{L_{\rm \nu}}{L_{\rm \nu, min}} \right)^{(\gamma + 1)} \, = N_{0} \, L_{\nu}^{\gamma + 1} = N_{0} \, L_{\rm \nu}^{\beta}  ,
\label{eq:singlepow_ccdf} 
\end{equation} where $N_{\rm tot} = \int^{\infty}_{L_{\rm \nu, min}} \, \rho(L_{\nu}) \, dL_{\nu}$ is the total number of pulsars and $N ( \geq L_{\nu}) = \int^{\infty}_{L_{\nu}} \, \rho(L_{\nu}) \, dL_{\nu}$, $N_0 = N_{\rm tot} \, L_{\rm \nu, min}^{-(\gamma + 1)}$, and $\beta = \gamma + 1$. It is clear from Eq. (\ref{eq:singlepow_ccdf}) that $N ( \geq L_{\nu})  = N_0 $ for $L_{\nu} = 1~{\rm mJy~kpc^2}$. Now I discuss the studies where people fitted Eq. (\ref{eq:singlepow_ccdf}) to pulsar luminosities. 

Allakhverdiev, Guseinov, and Tagieva \cite{agt97} fitted the CCFDF of $L_{400}$ with $\beta = -0.9 \pm 0.1$ and $N_{0} = 562$ for 68 pulsars with $L_{400} > 0.3~{\rm mJy~kpc^2}$.
Guseinov \textit{et~al.}\cite{gyto03} tried to find accurate luminosity functions for both $L_{400}$ and $L_{1400}$ using only pulsars closer than 1.5 kpc to obtain a sample with large enough numbers of low luminosity pulsars. They fitted broken power-laws for both $L_{400}$ and $L_{1400}$. For $L_{400}$, they found

\begin{equation}
\label{eq:brokenpowerlaw}
N(\geq L_{400})   = \left\{ \begin{array}{rl}
 520 \, L_{400}^{-0.85 \pm 0.01} &\mbox{~if $\log L_{400} > 1.5$ } \\
 62 \, L_{400}^{-0.19 \pm 0.01}  & \mbox{~if $0.2 < \log L_{400} < 1.5$ } \\
57.5 \, L_{400}^{-0.071 \pm 0.006}  & \mbox{~if $-0.5 < \log L_{400} < 0.2$   ,} \\
       \end{array} \right.
\end{equation}

and for $L_{1400}$, they found

\begin{equation}
\label{guseinov1400all}
N(\geq L_{1400})   = \left\{ \begin{array}{rl}
  188.4 \, L_{1400}^{-0.95 \pm 0.02} &\mbox{~if $\log L_{1400} > 0.5$ } \\
85.1 \, L_{1400}^{-0.27 \pm 0.01}  & \mbox{~if $-0.5 < \log L_{1400} < 0.5$} \\
 101 \, L_{1400}^{-0.13 \pm 0.01}    & \mbox{~if $-1.0 < \log L_{1400} < -0.5$ .} \\
       \end{array} \right.
\end{equation}

It is clear that in the above fits, the low luminosity part was much flatter for $L_{400}$ than that for $L_{1400}$. The authors suggested that the incompleteness of the pulsar surveys at $400$ MHz was responsible for this result. They also obtained luminosity functions for only isolated pulsars with characteristic ages $< 10^{7}$yr as

\begin{equation}
\label{guseinov400singleall}
N(\geq L_{400})   = \left\{ \begin{array}{rl}
  158.5 \, L_{400}^{-0.65 \pm 0.02} & \mbox{~if $\log L_{400} > 1.5$ } \\ 
 52.5 \, L_{400}^{-0.31 \pm 0.01} & \mbox{~if $0.4 < \log L_{400} < 1.5$~,}  
        \end{array} \right.
\end{equation} and

\begin{equation}
\label{guseinov1400singleall}
N(\geq L_{1400})   = \begin{array}{rl}
  66 \, L_{1400}^{-0.72 \pm 0.10}  & \mbox{~if $\log L_{1400} > 0.3$ \, .} 
       \end{array} 
\end{equation}

With a larger dataset (412 pulsars), an improved model to estimate the distances of the pulsars, and a Monte-Carlo method, Lorimer \textit{et~al.}\cite{lbdh93} found that the distribution of total galactic population of pulsars with $L_{400} > 10~{\rm mJy~kpc^2}$ could be expressed as $N(\geq L_{400}) = (7.34 \pm 1.06) \times 10^4 \, L_{400}^{-1}$ and the distribution of potentially observable pulsars with $L_{400} > 10~{\rm mJy~kpc^2}$ could be expressed as $N(\geq L_{400}) = (1.31 \pm 0.17) \times 10^4 \, L_{400}^{-1}$.

Using the flux densities and distance estimates available in the ATNF catalogue in 2010 for 51 MSPs ($P_s < 20$ ms) in the Galactic disk, Hui, Cheng and Taam\cite{hct10} fitted the CCFDF with Eq. (\ref{eq:singlepow_ccdf}) for different subcategories, i.e. the isolated, binary, and total population. I quote their best fit parameters in Table \ref{tab:hct10diskMSP}. They assumed $\sqrt{N ( \geq L_{1400})}$ as the uncertainties in the data (Poisson noise), and fitted $\log N ( \geq L_{1400})$ with a linear regression analysis. The power law became steeper when they kept only pulsars $L_{1400} \geq 1.5~{\rm mJy~ kpc^2}$, except for the case of isolated pulsars (which had a very small sample size).

\begin{table*}[h]
\tbl{Power law fit of $L_{1400}$ of MSPs in the Galactic disk by Hui, Cheng and Taam (2010).}
{\begin{tabular}{ |l|c|c|c|c|}
\hline
\multicolumn{2}{ |c| }{Pulsar Specification} & Sample Size &\multicolumn{2}{ |c| }{Fitting Parameters}  \\
\cline{4-5} 
 \multicolumn{2}{ |c| }{} &  & $N_0$ & $\beta$ \\ 
\hline
\multirow{2}{*}{Total} & all & 51 & $28^{+1}_{-1}$ & $-0.32 \pm 0.02$ \\
\cline{2-5}
  & only $L_{1400} \geq 1.5~{\rm mJy~ kpc^2}$ & 40 & $31^{+1}_{-1}$ & $- 0.48 \pm 0.04 $  \\ \hline
\multirow{2}{*}{Binary} & all & 39 &$25^{+1}_{-1}$ & $-0.36 \pm 0.03$ \\
\cline{2-5}
  & only $ L_{1400} \geq 1.5~{\rm mJy~ kpc^2}$  & 34  &$27^{+1}_{-1}$ & $- 0.49 \pm 0.05$ \\ \hline
\multirow{2}{*}{Isolated} & all & 12 & $5^{+1}_{-1}$ & $- 0.29 \pm 0.07$ \\
\cline{2-5}
  & only $ L_{1400} \geq 1.5~{\rm mJy~ kpc^2}$ & 6 & $4^{+2}_{-1}$ &  $-0.24 \pm 0.11 $ \\
\hline
\end{tabular}
\label{tab:hct10diskMSP}}
\end{table*}

I re-perform the analysis with the $L_{1400}$ values available in the ATNF catalogue in March 2013 for MSPs ($P_s < 20$ ms) in the Galactic disk, and the best fit parameters are given in Table \ref{tab:mnj13diskMSP}. Note that although the fitting coefficients are different from what obtained by Hui, Cheng and Taam, the fact that the power law become steeper when only pulsars with $L_{1400} \geq 1.5~{\rm mJy~ kpc^2}$ are considered, remains the same. I also fit the distribution of $L_{1400}$ keeping mildly recycled pulsars, i.e. taking all the pulsars with $P_s < 100$ ms, and the fit parameters are given in Table \ref{tab:mnj13diskrecycled}. For both of the cases, the total population can be fitted better with a double power law, as:
\begin{equation}
\label{eq:brokenpowerlaw}
N ( \geq L_{1400})  = \left\{ \begin{array}{rl}
 N_{0 l} \, L_{1400}^{\beta_l} &\mbox{ if $ L_{1400} \leq L_{\rm break} $} \\
 N_{0 h} \, L_{1400}^{\beta_h} & \mbox{ otherwise.} 
       \end{array} \right.
\end{equation}

\begin{figure*}
 \begin{center}
\hskip -2cm \subfigure[Single power law fit for all pulsars.]{\label{subfig:obslum_diskrecyclesingle}\includegraphics[width=0.5\textwidth,angle=0]{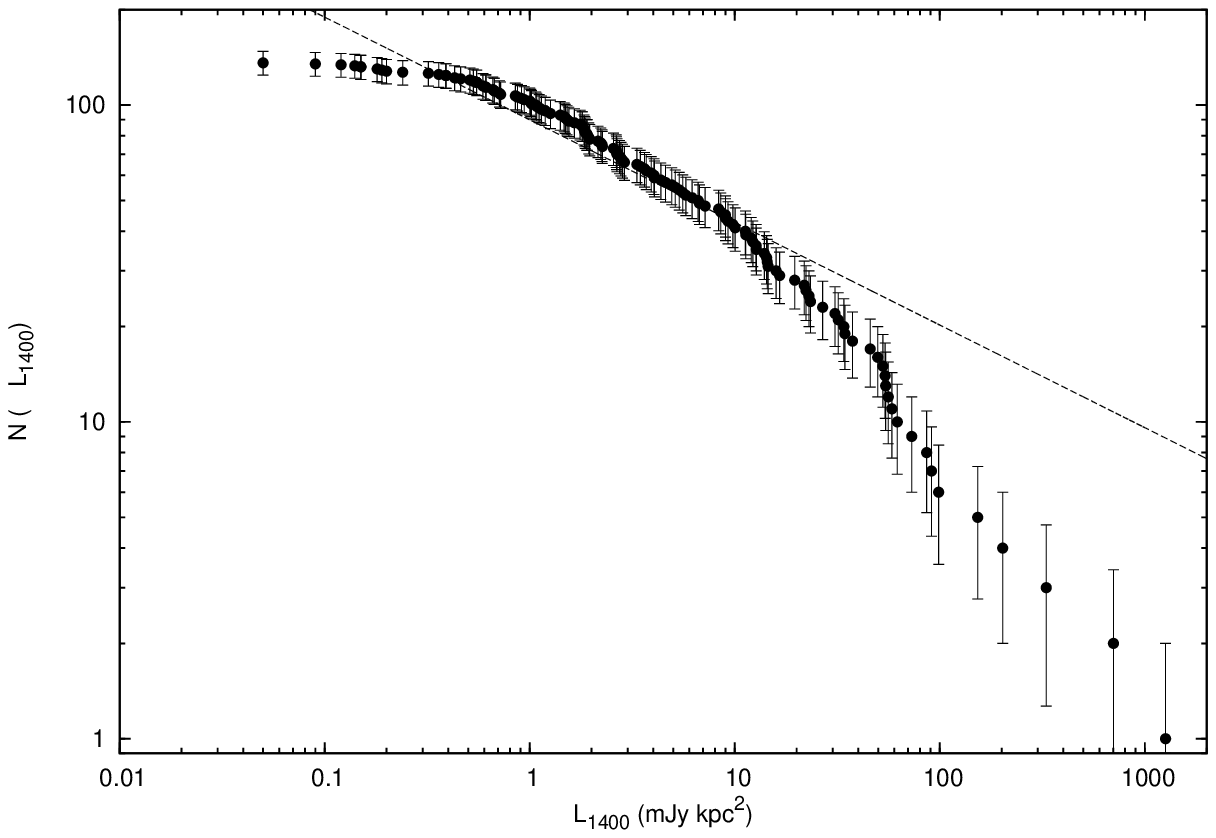}}
\hskip 1.0cm \subfigure[Double power law fit for all pulsars.]{\label{subfig:diskrecycledouble}\includegraphics[width=0.5\textwidth,angle=0]{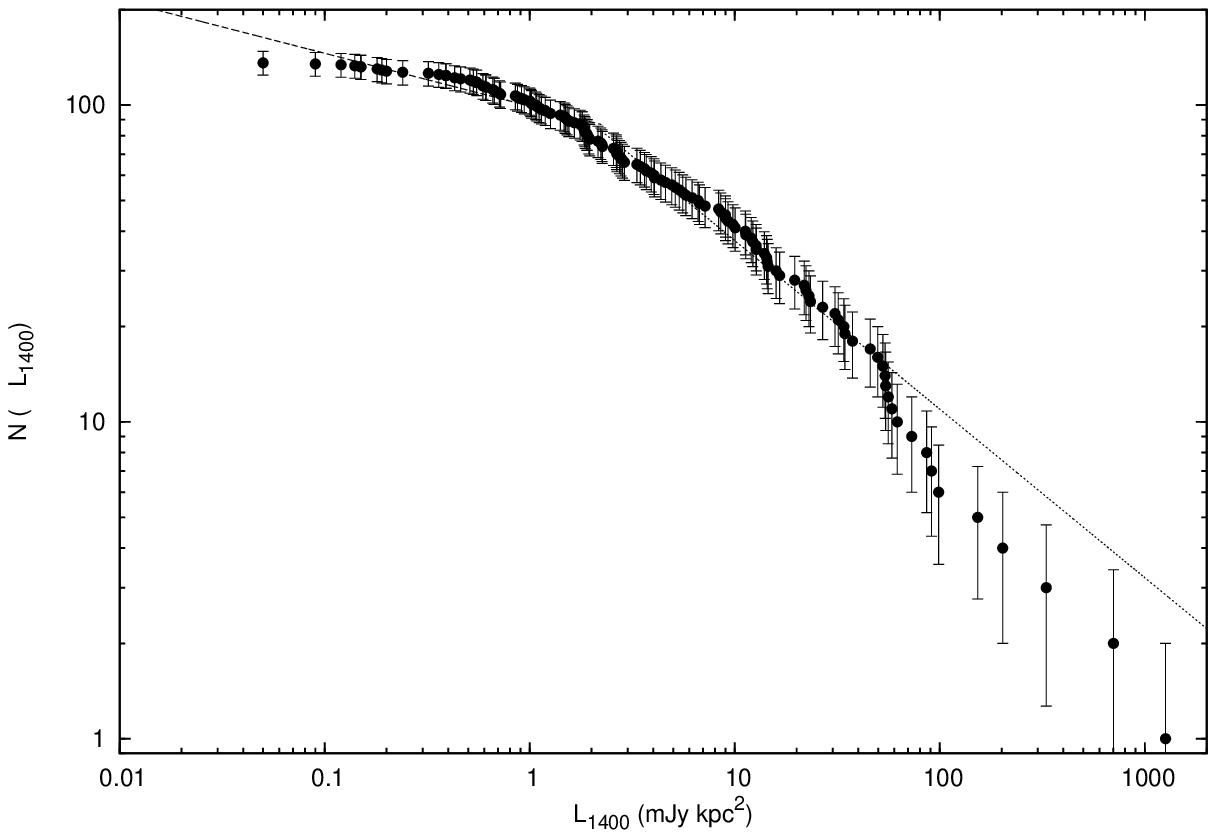}}
\vskip 0.1cm \hskip -2cm \subfigure[Single power law fit for pulsars with $L_{1400} \geq 1.5~{\rm mJy~kpc^2}$.]{\label{subfig:diskrecyclesinglebr}\includegraphics[width=0.5\textwidth,angle=0]{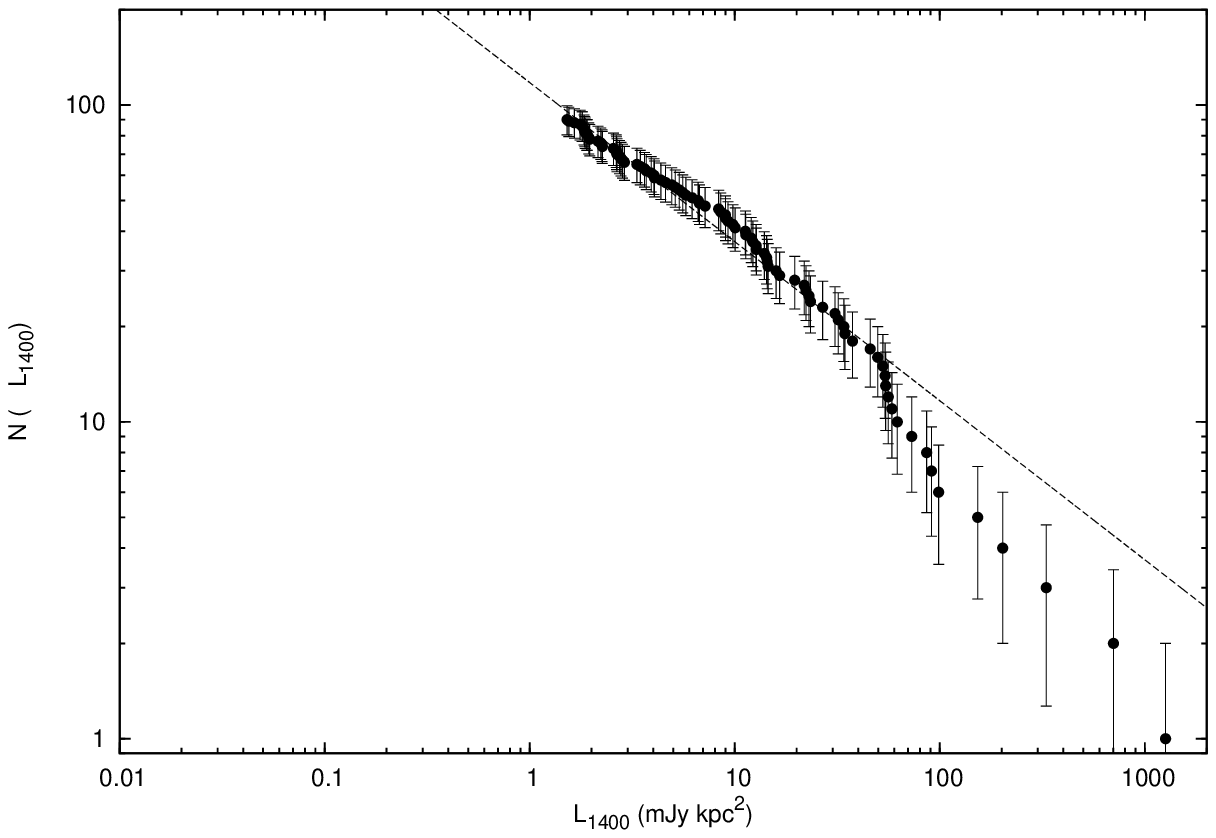}}
\hskip 1.0cm \subfigure[Double power law fit for pulsars with $L_{1400} \geq 1.5~{\rm mJy~kpc^2}$.]{\label{subfig:diskrecycledoublebr}\includegraphics[width=0.5\textwidth,angle=0]{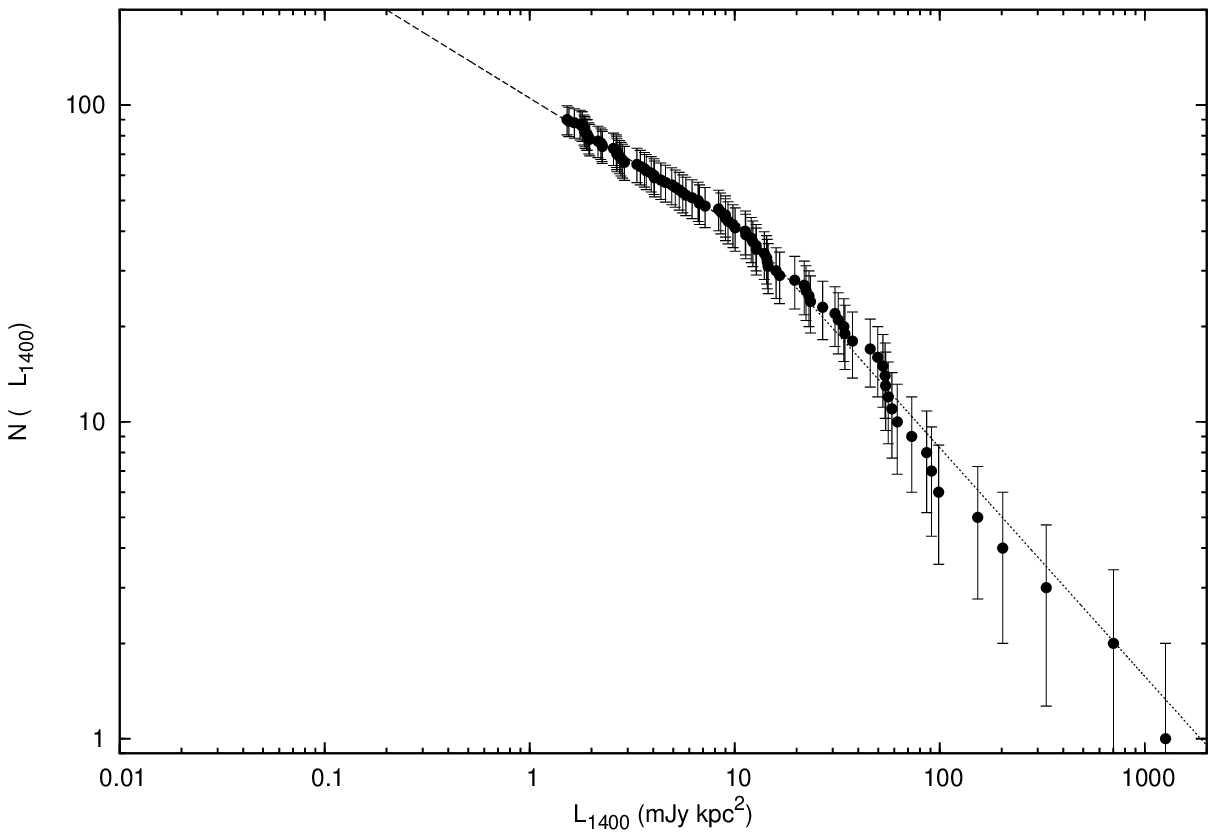}}
 \end{center}
\caption{Single and double power law fits for recycled pulsars ($P_s < 100$ ms) in the Galactic disk, the upper panel is for all pulsars and the lower panel is only for pulsars with $L_{1400} \geq 1.5~{\rm mJy~kpc^2}$. }
\label{fig:diskrecycledpowerlawfit}
\end{figure*}

The best fit parameters are $L_{\rm break} = 2.24 ~{\rm mJy~kpc^2}$, $N_{0 l} = 99^{+1}_{-1}$, $\beta_l =   -0.17 \pm 0.01 $, $N_{0 h} = 127^{+6}_{-5}$, and $\beta_h =  -0.53 \pm 0.02  $ if I keep all pulsars (with $P_s < 100$ ms) and $L_{\rm break} = 10.0 ~{\rm mJy~kpc^2}$, $N_{0 l} = 105^{+4}_{-4}$, $\beta_l =  -0.40  \pm 0.03 $, $N_{0 h} = 230^{+39}_{-33}$, and $\beta_h =  -0.72  \pm 0.05 $ if I keep only pulsars with $L_{1400} \geq 1.5~{\rm mJy~kpc^2}$ (and $P_s < 100$ ms). Fig. \ref{fig:diskrecycledpowerlawfit} shows the fits. For only short period pulsars ($P_s < 20$ ms), I get $L_{\rm break} = 1.17 ~{\rm mJy~kpc^2}$, $N_{0 l} = 59^{+2}_{-2}$, $\beta_l =  -0.15 \pm 0.03 $, $N_{0 h} = 61^{+3}_{-3}$, and $\beta_h =  -0.64 \pm 0.04  $ if I keep all pulsars and $L_{\rm break} = 12.59 ~{\rm mJy~kpc^2}$, $N_{0 l} = 59^{+4}_{-4}$, $\beta_l =  -0.59 \pm 0.06  $, $N_{0 h} = 98^{+65}_{-39}$, and $\beta_h =   -0.81 \pm 0.17  $ if I keep only pulsars with $L_{1400} \geq 1.5~{\rm mJy~kpc^2}$.

\begin{table*}[h]
\tbl{Power law fit of $L_{1400}$ of MSPs ($Ps < 20$ ms) in the Galactic disk by the author in 2013.}
{\begin{tabular}{ |l|c|c|c|c|}
\hline
\multicolumn{2}{ |c| }{Pulsar Specification} & Sample Size &\multicolumn{2}{ |c| }{Fitting Parameters}  \\
\cline{4-5} 
 \multicolumn{2}{ |c| }{} &  & $N_0$ & $\beta$ \\ 
\hline
\multirow{2}{*}{Total} & all & 82 & $47^{+1}_{-1}$ & $ -0.36 \pm 0.01  $ \\
\cline{2-5}
  & only $L_{1400} \geq 1.5~{\rm mJy~ kpc^2}$ & 46 & $62^{+4}_{-3}$ & $-0.65 \pm  0.04  $  \\ \hline
\multirow{2}{*}{Binary} & all & 62 &$37^{+1}_{-1}$ & $ -0.39  \pm 0.02  $ \\
\cline{2-5}
  & only $ L_{1400} \geq 1.5~{\rm mJy~ kpc^2}$  &  36 &$49^{+4}_{-3}$ & $ -0.68 \pm  0.05  $ \\ \hline
\multirow{2}{*}{Isolated} & all & 20 & $11^{+1}_{-1}$ & $ -0.30  \pm 0.04 $ \\
\cline{2-5}
  & only $ L_{1400} \geq 1.5~{\rm mJy~ kpc^2}$ & 10 & $13^{+3}_{-3}$ &  $-0.45 \pm  0.13   $ \\
\hline
\end{tabular}
\label{tab:mnj13diskMSP}}
\end{table*}

\begin{table*}[h]
\tbl{Power law fit of $L_{1400}$ of recycled pulsars ($Ps < 100$ ms) in the Galactic disk by the author in 2013.}
{\begin{tabular}{ |l|c|c|c|c|}
\hline
\multicolumn{2}{ |c| }{Pulsar Specification} & Sample Size &\multicolumn{2}{ |c| }{Fitting Parameters}  \\
\cline{4-5} 
 \multicolumn{2}{ |c| }{} &  & $N_0$ & $\beta$ \\ 
\hline
\multirow{2}{*}{Total} & all & 136 & $90^{+1}_{-1}$ & $- 0.32 \pm 0.01    $ \\
\cline{2-5}
  & only $L_{1400} \geq 1.5~{\rm mJy~ kpc^2}$ & 90 & $118^{+3}_{-3}$ & $ -0.50  \pm  0.02 $  \\ \hline
\multirow{2}{*}{Binary} & all & 83 &$53^{+1}_{-1}$ & $-0.38 \pm 0.01 $ \\
\cline{2-5}
  & only $ L_{1400} \geq 1.5~{\rm mJy~ kpc^2}$  & 52  &$70^{+4}_{-3}$ & $- 0.62 \pm  0.03   $ \\ \hline
\multirow{2}{*}{Isolated} & all & 53 & $38^{+1}_{-1}$ & $- 0.25 \pm 0.01   $ \\
\cline{2-5}
  & only $ L_{1400} \geq 1.5~{\rm mJy~ kpc^2}$ & 38 & $52^{+4}_{-4}$ &  $ -0.40 \pm  0.03 $ \\
\hline
\end{tabular}
\label{tab:mnj13diskrecycled}}
\end{table*}


\subsection{Population synthesis method}
\label{sublab:popsynth}

Faucher-Gigu\`ere and Kaspi\cite{fk06} performed a detailed population synthesis study of isolated, normal pulsars in the Galactic disk (the basic scheme was the same as what has been described in the antepenultimate paragraph of Section \ref{sec:lumlawppdot} of this article). They used two different luminosity laws for $L_{1400}$. The first one was a broken power law, independent of pulsar parameters, as follows:

\begin{equation}
\label{eq:fk06a}
\rho(L_{1400})  = \left\{ \begin{array}{rl}
 L_{1400}^{\gamma_1} &\mbox{ if $ L_{1400} \in [L_{1400}^{\rm low}, L_{1400}^{to})$} \\
 L_{1400}^{\gamma_2} &\mbox{ if $ L_{1400} \in [L_{1400}^{\rm to}, \infty)$} \\
  0 &\mbox{ otherwise,}
       \end{array} \right.
\end{equation} with parameters $L_{1400}^{\rm low} = 0.1~{\rm mJy~kpc^2}$, $L_{1400}^{\rm to} = 2.0 ~{\rm mJy~kpc^2}$, $\gamma_1 = - 1.267$, and $\gamma_2 = -2.0$. The second one was the conventional $P_{s} - \dot{P}_{s}$ law as:
\begin{equation}
\label{eq:fk06b}
L_{1400} = 10^{L_{1400, corr}} \, A_{0} \, P_{s}^{p} \, \dot{P}_{s}^{q} \, ,
\end{equation} where $L_{1400, corr}$ was chosen from a zero-centered normal distribution with a standard deviation of $\sigma_{L_{1400, corr}}$. The best fit values of the parameters were $\sigma_{L_{1400, corr}} = 0.8$, $A_{0} = 0.18~{\rm mJy~ kpc^2}$, $p=-1.5$, and $q=0.5$. By comparing the parameters of simulated (and selected) pulsars with the real ones, they noticed that the broken power law (Eq. \ref{eq:fk06a}) did not work well, specifically, it lead to too high values for the scale height (perpendicular distance from the Galactic disk) and produced a large number of pulsars near the ``death line" in the $P_{s} - \dot{P}_{s}$ diagram. On the other hand, the $P_{s} - \dot{P}_{s}$ dependent luminosity law (Eq. \ref{eq:fk06b}) worked fine, but interestingly, they noticed that the simulated distribution of $L_{1400}$ using this model could be simplified by a base-10 lognormal distribution, for which the PDF can be written as:
\begin{equation}
f_{\rm lognormal} \, (L_{1400})  =  \frac{\log_{10}e}{L_{1400}} \, \frac{1}{\sqrt{2 \pi \sigma^2}} \, \exp\left[{\frac{-(\log_{10} L_{1400} - \mu)^2}{2 \sigma^2}}\right].
\label{eq:lognorm_dist_def}
\end{equation} Here $\mu$ is the mean of the distribution of $\log_{10} L_{1400}$ and $\sigma$ is the standard deviation. The best fit values of these parameters were $\mu = -1.1$ and $\sigma = 0.9$, which translated to the mean of the distribution of $L_{1400}$ as $10^{-1.1}~{\rm mJy~kpc^2}$ or $0.079~{\rm mJy~kpc^2}$. Ridley and Lorimer (2010)\cite{rl10} performed almost similar study with much exhaustive models for the spin evolution of the pulsars and arrived at the same conclusion. 

Afterwards, Bagchi \textit{et~al.}\cite{manjari11} concluded that this lognormal luminosity model worked fine for luminosities of recycled pulsars in the Galactic globular clusters, although they could not constrain the parameter space very well, which was later tried by Chennamangalam \textit{et~al.} (2013)\cite{jayanth13}. These two works will be discussed in details in the next section.


\section{Luminosity function for pulsars in globular clusters}
\label{lab:GCpulsar}

For pulsars in globular clusters (GCs), it is difficult to model the effects of stellar encounters and the cluster potential\cite{phinney92}. So the population synthesis method to study the luminosity distribution of cluster pulsars is not popular, the direct method is commonly used.

Anderson\cite{and1992} fitted $L_{430}$ of 8 pulsars in the GC M31 using Eqn. (\ref{eq:powerlaw1}) with $\gamma = -2.00 \pm 0.35$ and 6 isolated pulsars with $\gamma = -1.93 \pm 0. 0.38$. Kulkarni \textit{et~al.}\cite{kgw90} performed interferometric studies using the Very Large Array (VLA) and measured total diffuse radio fluxes at 1400 MHz for few GCs as M 4, M 28, M 15 and M 13, as well as one point source in each of these clusters. They used the CCFDF as defined in Eq. (\ref{eq:singlepow_ccdf}) with $\gamma = -2$ and $L_{\rm 1400, min} = 1.0~{\rm mJy \, kpc^2}$ to predict the total number of pulsars in these clusters, as well as tried to take care of cluster properties with a simplified manner, by introducing a weight factor. At the same time, Fruchter and Goss\cite{fg90} performed an almost similar study for few other GCs using $\gamma = - 1.85$ and $L_{\rm 1400, min} = 0.2~{\rm mJy \, kpc^2}$. This choice of $L_{\rm 1400, min}$ was based on the fact that the minimum value of $L_{1400}$ known at that time was around $0.2~{\rm mJy \, kpc^2}$ and the minimum value of $L_{400}$ known at that time was around $3~{\rm mJy \, kpc^2}$ which gave the value of $L_{1400}$ between $0.31$ to $0.24~{\rm mJy \, kpc^2}$ for the spectral index in the range of $-1.8$ to $-2.0$. By equating the sum of the simulated (following that power law) flux values, to the total diffuse flux, they concluded that Terzan 5 contained about 75 potentially observable pulsars, NGC 6440 around 60 and NGC around 20, and total number of observable pulsars in all Galactic globular clusters were in the range of $500 - 1800$. Afterwards, they reperformed\cite{fg00} the analysis with improved data using both VLA and ATCA (Australia Telescope Compact Array), where they measured total diffuse radio fluxes at 1400 MHz for few GCs, as well as fluxes of few point sources (pulsars) for Terzan 5 and 47 Tuc. They used a power law with $\gamma = - 1.85$ and $L_{\rm 1400, min} = 0.3~{\rm mJy \, kpc^2}$ to fit the brightest point sources in their data. They predicted that the total number of pulsars in Terzan 5 was between $60 - 200$.  Later, McConnell and Deshpande\cite{mdca04} fitted the $S_{1400}$ of the pulsars in 47 Tuc (as reported by Camilo \textit{et~al.}\cite{clflm00}) with a power law. More specifically, they fitted the CCFDF of $S_{1400}$ as $N ( \geq S_{1400}) = 10^{-0.1 \pm 0.2} \, S_{1400}^{-0.9 \pm 0.2}$ when both the parameters were free, and as $N ( \geq S_{1400}) = 10^{-0.18 \pm 0.07} \, S_{1400}^{-1}$ when they fixed the power law index as $-1$. They came to the conclusion that the upper limit of the number of observable pulsars in 47 Tuc was around 30.

Similar to their analysis for MSPs in the Galactic disk, Hui, Cheng and Taam\cite{hct10} fitted a power law (Eq. \ref{eq:singlepow_ccdf}) for GC pulsars. They selected only such GC pulsars, for which each GC had atleast four pulsars with published values of flux densities. They used a spectral index of $-1.8$ to convert flux densities measured at other frequencies to 1400 MHz and the distances of globular clusters published in Harris catalogue (http://physwww.physics.mcmaster.ca/$\sim$harris/mwgc.dat)\cite{harris96} to obtain $L_{1400}$. I quote their best fit parameters in Table \ref{tab:hct10gcMSP}. The power law became steeper when only pulsars with $L_{1400} \geq 1.5~{\rm mJy~ kpc^2}$ were considered. I also compare these fit parameters with the ones obtained by Hessels \textit{et~al.}\cite{hrskf07} (only $\beta$, values of $N_0$ were not reported by Hessels \textit{et~al.}). 

Hui, Cheng and Taam\cite{hct10} also concluded that the luminosities of MSPs in GCs were different from those in the Galactic disk as the CCFDF for GC MSPs was much steeper than that of disk pulsars (by comparing Table \ref{tab:hct10diskMSP} with Table \ref{tab:hct10gcMSP}). This is a very important conclusion. If correct, it would imply that the radio luminosity is related to differences in formation processes between the disk and GC pulsars. The same analysis was re-performed by Bagchi and Lorimer\cite{bl10} with more recent distance estimates of GCs and the resultant CCFDF was even steeper. They obtained $N_0 = 59^{+1}_{-1}$ and $\beta=-0.80 \pm 0.03$ when they kept all pulsars and $N_0 = 74^{+5}_{-4}$ and $\beta=-1.06 \pm 0.06$ when they kept only the pulsars with $L_{1400} \geq 1.5~{\rm mJy~ kpc^2}$. Remember that although for disk pulsars, only pulsars with $P_s < 20$ ms were chosen, no such criterion was chosen for GC pulsars by Hui, Cheng and Taam; and Bagchi and Lorimer used a condition $P_s < 100$ ms. Hui, Cheng and Taam also fitted power laws for pulsars in different GCs separately keeping only pulsars with $L_{1400} \geq 0.5~{\rm mJy~ kpc^2}$, which was also redone by Bagchi and Lorimer. The comparison is in Table \ref{tab:simpleGCcompare}. Again, we need to remember the difference between two datasets. Hui, Cheng and Taam did not exclude pulsars with $P_{s} \geq 100$ ms, which Bagchi and Lorimer did. Moreover, Bagchi and Lorimer used latest distance estimates for GCs to convert $S_{1400}$ to $L_{1400}$. The largest discrepancy in the value of the distance was for the case of Terzan 5, for which Hui, Cheng and Taam chose $d=10.3$ kpc but Bagchi and Lorimer chose $d=5.5$ kpc. Bagchi and Lorimer also showed that a double power law (Eq \ref{eq:brokenpowerlaw}) fitted the luminosities of recycled pulsars in GCs better than a single power law. The best fit values of the parameters were $L_{\rm break} = 4.0~{\rm mJy~kpc^2}$, $N_{0 l} = 70^{+7}_{-6}$, $\beta_l = -0.97 \pm 0.13$, $N_{0 h} = 134^{+60}_{-41}$, and $\beta_h = -1.40 \pm 0.21$, when they kept only pulsars with $L_{1400} \geq 1.5~{\rm mJy~ kpc^2}$.

\begin{table*}[h]
\tbl{Power law fit of $L_{1400}$ of MSPs in the Galactic globular clusters, as obtained by Hui, Cheng and Taam (2010) and Hessels \textit{et~al.} (2007).}
{\begin{tabular}{ |l|c|c|c|c|c|c c|}
\hline
\multicolumn{2}{ |c| }{} &  \multicolumn{6}{ |c| }{Fitting Parameters}  \\
\cline{3-8} 
\multicolumn{2}{ |c| }{Pulsar Specification} &  \multicolumn{3}{ |c }{Hui, Cheng and Taam (2010)} &\multicolumn{3}{ |c| }{Hessels \textit{et~al.} (2007)} \\
\cline{3-8} 
 \multicolumn{2}{ |c| }{} & Sample Size & $N_0$ & $\beta$ &  Sample Size & \multicolumn{2}{ |c| }{ $\beta$ } \\ 
\hline
\multirow{2}{*}{Total} & all & 76 & $68^{+2}_{-2}$ & $-0.58 \pm 0.03$ & 82 &\multicolumn{2}{ |c| }{-} \\
\cline{2-8}
  & only $L_{1400} \geq 1.5~{\rm mJy~ kpc^2}$ & 58 & $91^{+7}_{-6}$ & $-0.83 \pm 0.05$  & 70 &\multicolumn{2}{ |c| }{$- 0.77\pm 0.03$}\\ \hline
\multirow{2}{*}{Binary} & all & 41 &$36^{+2}_{-2}$ & $-0.56 \pm 0.05$ & 41& \multicolumn{2}{ |c| }{-} \\
\cline{2-8}
  & only $ L_{1400} \geq 1.5~{\rm mJy~ kpc^2}$  &  32 &$44^{+4}_{-4}$ & $-0.73 \pm 0.08$ & 33 & \multicolumn{2}{ |c| }{$-0.63 \pm 0.06$} \\ \hline
\multirow{2}{*}{Isolated} & all & 33 & $32^{+2}_{-2}$ & $-0.61 \pm 0.06$ & 41 & \multicolumn{2}{ |c| }{-} \\
\cline{2-8}
  & only $ L_{1400} \geq 1.5~{\rm mJy~ kpc^2}$ & 26 & $47^{+7}_{-6}$ &  $-0.89 \pm 0.11$ & 37 & \multicolumn{2}{ |c| }{$-0.90 \pm 0.07$}\\
\hline
\end{tabular}
\label{tab:hct10gcMSP}}
\end{table*}

\begin{table}[h]
\tbl{Power law fit parameters for pulsars in different globular clusters, keeping only pulsars with only pulsars with $L_{1400} \geq 0.5~{\rm mJy~ kpc^2}$.}
{\begin{tabular}{|l|c|c|c|c|} \hline
{GC  Name} & \multicolumn{2}{ |c| }{Hui, Cheng, and Taam\cite{hct10}}  &\multicolumn{2}{ |c| }{ Bagchi and Lorimer\cite{bl10}}  \\
\cline{2-5}
& $N_0$ & $\beta$ & $N_0$ & $\beta$ \\ \hline
47Tuc & $11^{+2}_{-2}$    & $ -0.82 \pm 0.19$ & $10^{+1}_{-1}$ & $ -0.85 \pm 0.18 $   \\
M3 & $2^{+1}_{-1}$  & $ -1.61 \pm 1.09 $ & $2^{+1}_{-1}$ & $ -1.52 \pm 1.14$  \\
M5 & $3^{+1}_{-1}$  & $ -0.58 \pm 0.38$ & $3^{+1}_{-1}$ & $ -0.55 \pm 0.32$  \\
M13 & $4^{+2}_{-1}$  & $ -0.63 \pm 0.34$ & $4^{+1}_{-1}$ & $ -0.62 \pm 0.39 $  \\
Ter5 & $50^{+12}_{-9}$  & $ -0.80 \pm 0.12 $ & $20^{+1}_{-1}$ & $ -0.87 \pm 0.10$  \\
NGC 6440 & $10^{+7}_{-4}$  & $ -0.59 \pm 0.27 $ & $11^{+12}_{-6}$ & $ 0.86 \pm 0.53$  \\
NGC 6441 & $8^{+14}_{-5}$  & $ -0.76 \pm 0.52$ & -- & --  \\
M28 & $10^{+5}_{-4}$  & $-0.74 \pm 0.26 $ & $12^{+4}_{-3}$ & $ 0.91 \pm 0.31$  \\
NGC 6752 & $5^{+2}_{-2}$  & $ -0.93 \pm 0.50 $ & $5^{+2}_{-1}$ & $ -0.78 \pm 0.44 $  \\
M15 & --  & -- &  $8^{+3}_{-2}$& $ -0.83 \pm 0.34 $  \\
\hline
\end{tabular}}
\label{tab:simpleGCcompare}
\end{table}

\begin{figure*}
 \begin{center}
\hskip -2cm \subfigure[Single power law fit for all pulsars.]{\label{subfig:gcrecycledsingle}\includegraphics[width=0.5\textwidth,angle=0]{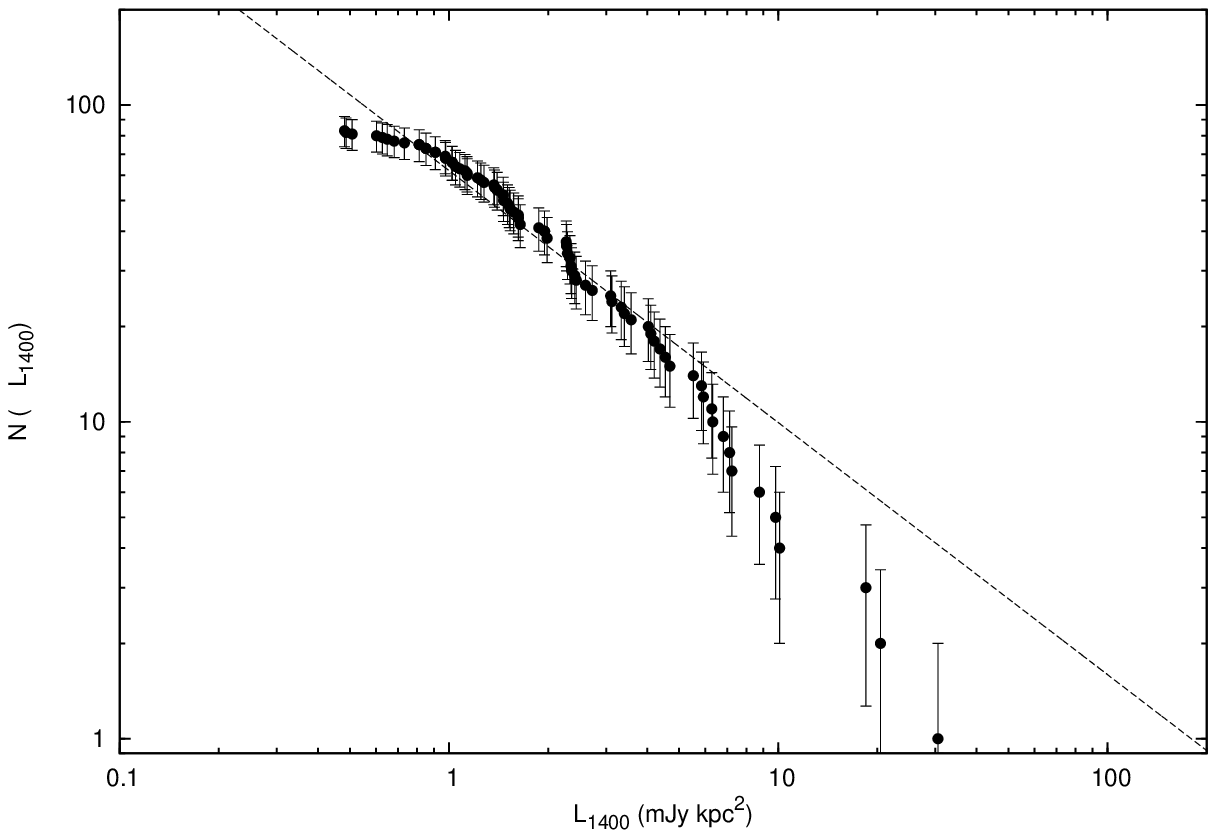}}
\hskip 1.0cm \subfigure[Double power law fit for all pulsars.]{\label{subfig:gcrecycleddouble}\includegraphics[width=0.5\textwidth,angle=0]{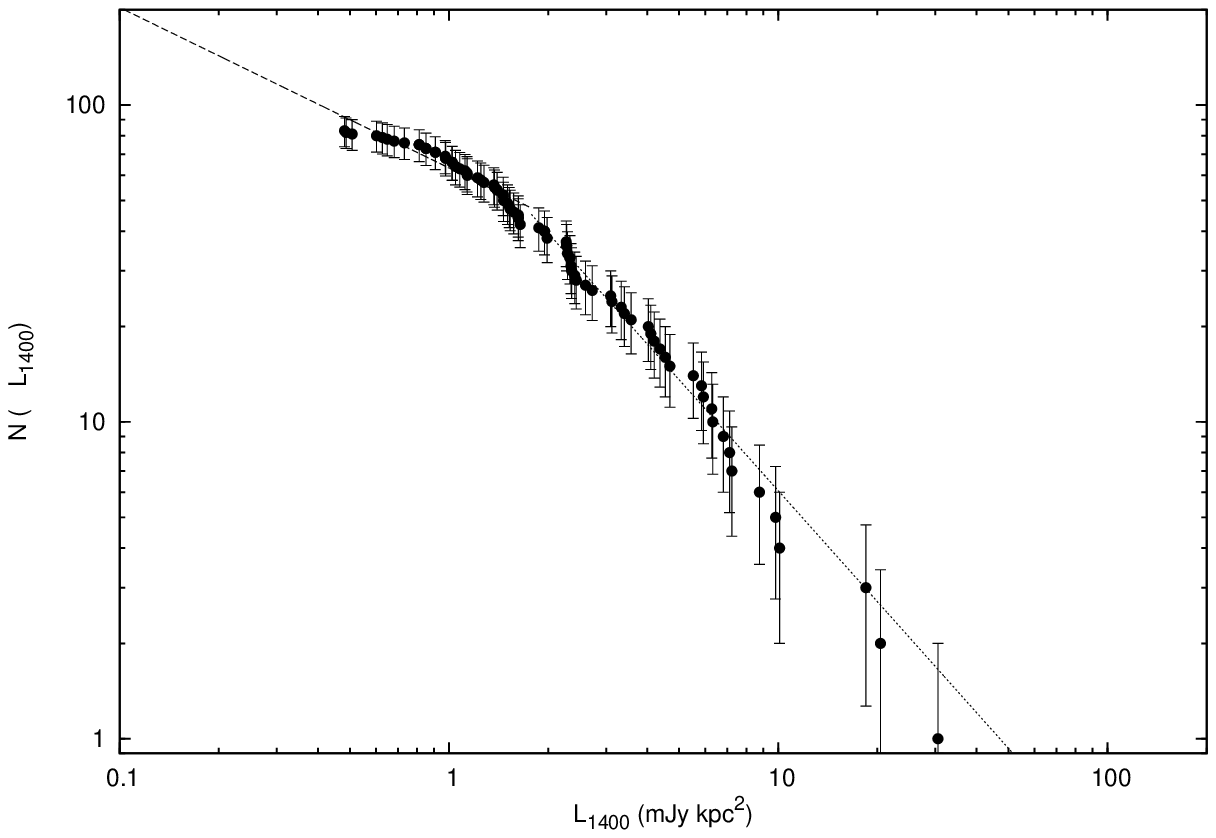}}
\vskip 0.1cm \hskip -2cm \subfigure[Single power law fit for pulsars with $L_{1400} \geq 1.5~{\rm mJy~kpc^2}$.]{\label{subfig:gcrecycledsinglebr}\includegraphics[width=0.5\textwidth,angle=0]{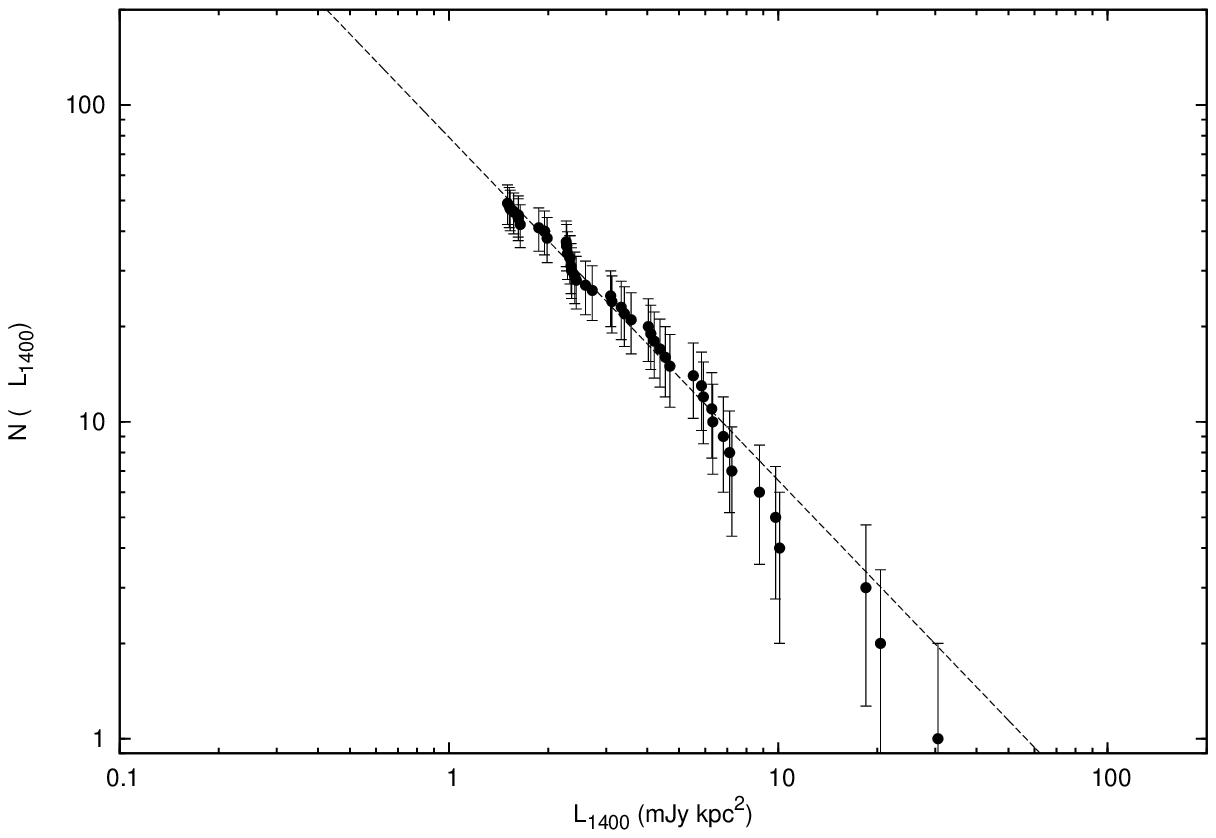}}
\hskip 1.0cm \subfigure[Double power law fit for pulsars with $L_{1400} \geq 1.5~{\rm mJy~kpc^2}$.]{\label{subfig:gcrecycleddoublebr}\includegraphics[width=0.5\textwidth,angle=0]{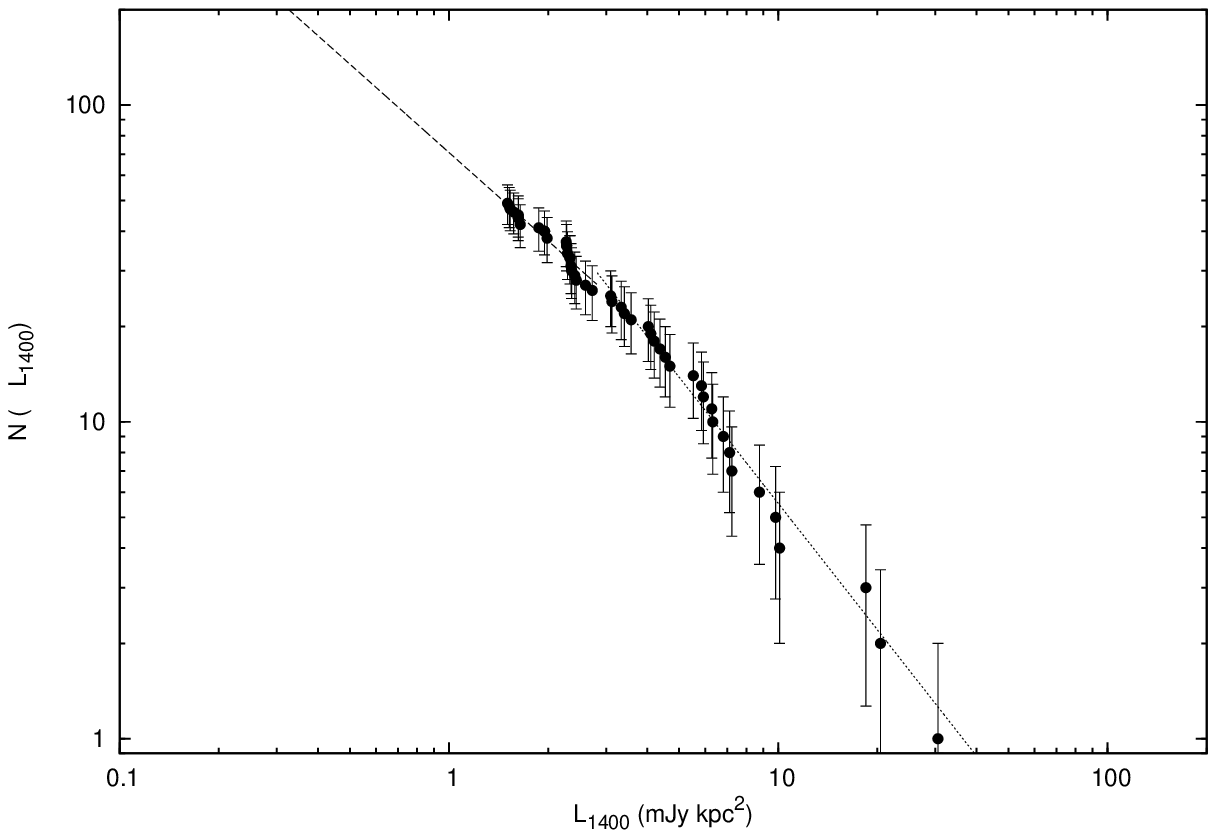}}
 \end{center}
\caption{Single and double power law fits for recycled pulsars ($P_s < 100$ ms) in GCs, the upper panel is for all pulsars and the lower panel is only for pulsars with $L_{1400} \geq 1.5~{\rm mJy~kpc^2}$. }
\label{fig:gcrecycledpowerlawfit}
\end{figure*}

At this time, flux estimates for the pulsars in NGC 6517 have been published and were included in the later study by Bagchi \textit{et~al.}\cite{manjari11}. Here, for the sake of comparison, I re-perform the analysis similar to Bagchi and Lorimer\cite{bl10} with this improved dataset. Interested readers can consult Tables 1 and 4 of Bagchi \textit{et~al.}\cite{manjari11} for details of pulsar parameters (including $P_{s}$, $S_{1400}$, cluster association etc.) and cluster parameters. The best fit parameters for pulsars with $P_s < 100$ ms are given in Table \ref{tab:mnj13gcMSP} and the best fit parameters for pulsars with $P_s < 20$ ms (which was the condition used for disk MSPs) are given in Table \ref{tab:mnj13gcMSPshort}. Table \ref{tab:mnj13gcMSPshort} (which shows the best fit parameters for GC MSPs) should be compared with Table \ref{tab:mnj13diskMSP} (which shows the best fit parameters for disk MSPs). For each category, i.e. the binary, isolated and total population, the luminosity distribution of MSPs in GCs is much steeper than that of the MSPs in the Galactic disk. A double power law (Eq. \ref{eq:brokenpowerlaw}) still gives a better fit than a single power law. The best fit parameters are $L_{\rm break} = 1.78 ~{\rm mJy~kpc^2}$, $N_{0 l} = 63^{+1}_{-1}$, $\beta_l =  -0.51 \pm 0.05$, $N_{0 h} = 88^{+9}_{-8}$, and $\beta_h =  -1.16 \pm 0.08 $ if I keep all pulsars (with $P_s < 100$ ms) and $L_{\rm break} = 2.82~{\rm mJy~kpc^2}$, $N_{0 l} = 71^{+9}_{-8}$, $\beta_l =  -0.93 \pm 0.18 $, $N_{0 h} = 117^{+31}_{-25}$, and $\beta_h =  -1.32  \pm 0.15 $ if I keep only pulsars with $L_{1400} \geq 1.5~{\rm mJy~kpc^2}$ (and $P_s < 100$ ms). Fig. \ref{fig:gcrecycledpowerlawfit} shows the fits. For only short period pulsars ($P_s < 20$ ms), I get $L_{\rm break} =  1.6~{\rm mJy~kpc^2}$, $N_{0 l} = 59^{+1}_{-1}$, $\beta_l =  -0.45 \pm 0.06 $, $N_{0 h} = 80^{+7}_{-6}$, and $\beta_h =   -1.16 \pm 0.08  $ if I keep all pulsars and are $L_{\rm break} = 2.8 ~{\rm mJy~kpc^2}$, $N_{0 l} = 61^{+7}_{-7}$, $\beta_l =   -0.81 \pm 0.16  $, $N_{0 h} = 104^{+30}_{-23}$, and $\beta_h =  -1.33 \pm 0.16  $ if I keep only pulsars with $L_{1400} \geq 1.5~{\rm mJy~kpc^2}$.

\begin{table*}[h]
\tbl{Power law fit of $L_{1400}$ of recycled pulsars ($P_s < 100$ ms) in the Galactic globular clusters, as obtained by the author in 2013.}
{\begin{tabular}{ |l|c|c|c|c|}
\hline
\multicolumn{2}{ |c| }{Pulsar Specification} & Sample Size &\multicolumn{2}{ |c| }{Fitting Parameters}  \\
\cline{4-5} 
 \multicolumn{2}{ |c| }{} &  & $N_0$ & $\beta$ \\ 
\hline
\multirow{2}{*}{Total} & all & 83 & $62^{+1}_{-1}$ & $-0.80 \pm 0.03$ \\
\cline{2-5}
  & only $L_{1400} \geq 1.5~{\rm mJy~ kpc^2}$ & 49 & $79^{+5}_{-5}$ & $-1.08 \pm 0.06 $  \\ \hline
\multirow{2}{*}{Binary} & all & 42 &$32^{+1}_{-1}$ & $-0.80 \pm 0.06$ \\
\cline{2-5}
  & only $ L_{1400} \geq 1.5~{\rm mJy~ kpc^2}$  & 25  &$37^{+5}_{-4}$ & $-1.03 \pm 0.12$ \\ \hline
\multirow{2}{*}{Isolated} & all & 39 & $31^{+1}_{-1}$ & $- 0.77 \pm 0.06 $ \\
\cline{2-5}
  & only $ L_{1400} \geq 1.5~{\rm mJy~ kpc^2}$ & 24 & $41^{+6}_{-5}$ &  $- 1.07 \pm 0.13 $ \\
\hline
\end{tabular}
\label{tab:mnj13gcMSP}}
\end{table*}

\begin{table*}[h]
\tbl{Power law fit of $L_{1400}$ of MSPs ($P_s < 20$ ms) in the Galactic globular clusters, as obtained by the author in 2013.}
{\begin{tabular}{ |l|c|c|c|c|}
\hline
\multicolumn{2}{ |c| }{Pulsar Specification} & Sample Size &\multicolumn{2}{ |c| }{Fitting Parameters}  \\
\cline{4-5} 
 \multicolumn{2}{ |c| }{} &  & $N_0$ & $\beta$ \\ 
\hline
\multirow{2}{*}{Total} & all & 76 & $57^{+1}_{-1}$ & $-0.78  \pm 0.03 $ \\
\cline{2-5}
  & only $L_{1400} \geq 1.5~{\rm mJy~ kpc^2}$ & 46 & $73^{+5}_{-5}$ & $- 1.08 \pm 0.07 $  \\ \hline
\multirow{2}{*}{Binary} & all & 38 &$29^{+1}_{-1}$ & $- 0.78  \pm 0.06$ \\
\cline{2-5}
  & only $ L_{1400} \geq 1.5~{\rm mJy~ kpc^2}$  & 23  &$35^{+5}_{-4}$ & $- 1.04 \pm 0.13 $ \\ \hline
\multirow{2}{*}{Isolated} & all & 36 & $28^{+1}_{-1}$ & $-0.76 \pm 0.06 $ \\
\cline{2-5}
  & only $ L_{1400} \geq 1.5~{\rm mJy~ kpc^2}$ & 22  & $41^{+7}_{-6}$ &  $-1.13 \pm 0.15 $ \\
\hline
\end{tabular}
\label{tab:mnj13gcMSPshort}}
\end{table*}

As we already know, the direct method to obtain the luminosity function does not take care of selection effects, and as GCs are in general at large distances, it is difficult to detect low luminosity pulsars in GCs, thus the observed sample is more biased towards bright pulsars, than that for the disk pulsars. To overcome this limitation, Bagchi \textit{et~al.}\cite{manjari11} introduced an improved method where they modelled the luminosity distribution of recycled pulsars in globular clusters as the brighter, observable part of an intrinsic (parent) distribution, assuming that the parent luminosity distribution is the same for all GCs. They considered the sample of 83 pulsars (with known flux values) with spin periods $P \leq 100$ ms in 10 GCs with the condition that each of these GCs hosted at least 4 such pulsars. For all these objects, the spin and binary properties suggested that the neutron stars had undergone the phase of recycling in the past. They converted flux densities measured at other frequencies to $S_{1400}$ using the estimated values of $\alpha$ from observed values of fluxes at different frequencies whenever available, otherwise $\alpha = -1.9$. Details of flux densities and spectral indices for GC pulsars can be found in Table 1 of Bagchi \textit{et~al}. They used latest distance estimates of GCs to obtain $L_{\nu}$ from $S_{\nu}$ (Eq. \ref{eq:pseudolum}). It is a well known fact that the measured values of $\dot{P}_{s}$ for GC pulsars are affected by cluster potentials \cite{phinney92}, and Bagchi \textit{et~al.}\cite{manjari11} did not find any correlation between $L_{1400}$ and $P_{s}$ for GC pulsars. That is why instead of using any $P_s$, $\dot{P}_{s}$ dependent luminosity law (as described in Section \ref{sec:lumlawppdot}), they used simple luminosity distribution functions. Using those distribution functions, they generated synthetic samples of pulsar luminosities in each GC and selected only those pulsars which had simulated luminisities greater than the minimum observed luminosity for that GC, until they obtained the desired number of such bright and selected pulsars in each case. The number of pulsars they needed to generate to obtain such desired number of bright pulsars gave estimates of total number of pulsars in each GC, and the sum of the fluxes of all these pulsars gave estimates of total fluxes from each GC. They combined luminosities of all such bright pulsars to get a combined luminosity distribution which they compared with the observed distribution. A key assumption in this method was that each GC had been searched down to the level of the faintest observable pulsar in that particular cluster. This assumption provided a good approximation to the actual survey sensitivity in each cluster, and was made primarily due to the lack of published detail of several of the globular cluster surveys.

The first distribution function chosen by Bagchi \textit{et~al.} was the lognormal luminosity function (see Eq. \ref{eq:lognorm_dist_def} for the PDF). They chose both $\mu$ and $\sigma$ as free parameters and perform their analysis (as mentioned in the above paragraph) for each set of values of $\mu$, $\sigma$. To compare the simulated luminosity distributions with the observed one, they performed two statistical tests, the KS test and the $\chi^2$ test. For both of the tests, they obtained wide ranges of values of $\mu$ and $\sigma$ providing good fits. For examples, they quoted that the KS test resulted the best fit (maximum value of $P_{\rm KS} = 0.98$) for $\mu = -0.61$ and $\sigma = 0.65$, while the $\chi^2$ test resulted the best fit (minimum value of $\chi^2 = 6.3$) for $\mu = -0.52$ and $\sigma = 0.62$. The 2-$\sigma$ contour around the minimum $\chi^2$ and the region of $P_{\rm KS} \geq 0.05$ were almost the same and enclosed large areas in the $\mu - \sigma$ space. The model used by Faucher-Gigu\`ere and Kaspi\cite{fk06} for isolated, slow pulsars in the Galactic disk ($\mu=-1.1$ and $\sigma=0.9$) fell inside those regions, i.e. provided a good fit. The next distribution they used was the power law\footnote{Remember $\beta = (\gamma+1)$ according to the notation used in this article, while Bagchi \textit{et~al.} used the notation $\beta = -(\gamma+1)$.}, for which the PDF is given in Eq (\ref{eq:singlepow_dist_def}). They chose $L_{\rm 1400, min}$ in the range of $0.003-0.48 ~{\rm mJy~kpc^2}$, as 0.48 ${\rm mJy~kpc^2}$ was the observed minimum luminosity among GC pulsars in their sample, and the lower value of $L_{\rm 1400, min}$ was chosen somewhat arbitrarily. They also chose a maximum luminosity of 50~mJy~kpc$^2$ and argued that there was no GC pulsar with $L_{1400}>20$~mJy~kpc$^2$, and their results were insensitive to the exact choice of the maximum luminosity cutoff over the range of 20--500~mJy~kpc$^2$. Here the KS test resulted the best fit (maximum value of $P_{\rm KS} = 0.81$) for $\gamma = -1.92$, $L_{\rm 1400, min} = 0.017 {\rm~ mJy~kpc^2}$ and the $\chi^2$ test resulted the best fit (minimum value of $\chi^2 = 8.0$) for $\gamma = -2.01$, $L_{\rm 1400, min} = 0.022{\rm~ mJy~kpc^2}$. Based on this, they concluded that the lognormal function was statistically slightly better description for the luminosity distribution of recycled pulsars in GCs. They also demonstrated that an exponential distribution with PDF $ f_{\rm exponential} \, (L_{1400})  =  \lambda e^{-\lambda L_{1400}} $ did not work well for any value of the parameter $\lambda$ ($1/{\lambda}$ is the mean of the distribution). They also predicted the total number of pulsars and total flux densities for their best fit models (see Tables 2 and 3 of Bagchi \textit{et~al.}). The best fit lognormal models predicted $60-200$ pulsars in Terzan 5 and $17 - 90$ pulsars in 47 Tuc, while the best fit power law models predicted $290 -  978$ pulsars in Terzan 5 and $ 112 - 411$ pulsars in 47 Tuc. The predicted numbers of pulsars were much larger for power law models, which predicted large numbers of faint pulsars. They also compared their simulated values of total flux densities to observed values of diffuse radio flux densities of Terzan 5 and 47 Tuc, assuming that the only contributions to these fluxes were from pulsars. For Terzan 5, they used the diffuse flux density measured by Fruchter and Goss\cite{fg00} which was $S_{\rm 1400, obs, tot} = 5.2 ~{\rm mJy~kpc^2}$ (sum of the diffuse flux and the fluxes of point sources, i.e. known pulsars at that time). For 47 Tuc, they used the diffuse flux density measured by McConnell and Deshpande\cite{mdca04} as $S_{\rm obs, tot} = 2.0 \pm 0.3 ~{\rm mJy~kpc^2}$. Bagchi \textit{et~al.} noticed that all their best models (mentioned earlier) could reproduce the observed diffuse flux for 47 Tuc. For Terzan~5, the power law models provided better matches to the diffuse flux overall, while the lognormal models predicted slightly smaller fluxes which lied 2--5$\sigma$ below the nominal value found by Fruchter and Goss\cite{fg00}. To constrain the parameter space further, first they worked on the lognormal distribution, where they fixed $\mu$ as $-1.1$ and varied $\sigma$. They noticed that there were only two possible ranges of $\sigma$ which were compatible with the diffuse flux from Terzan 5: $\sigma \sim 0.5$ or $\sigma \sim 0.9$, but they preferred the ``solution" with $\sigma \sim 0.9$ as the set $\mu = -1.1$, $\sigma = 0.5$ did not provide good $\chi^2$ fitting. Remember, $\mu = -1.1$, $\sigma = 0.9$ is the parameter set preferred by Faucher-Gigu\`ere and Kaspi for isolated, normal pulsars in the Galactic disk. Similarly, for the power law distribution, fixing $\gamma = -2$, they found that a wide range of values of $L_{\rm 1400, min}$ were consistent with the diffuse flux from Terzan 5. As the lower end of this range gave an unrealistically large number of pulsars, they preferred the upper end of this range, which corresponded to $L_{\rm 1400, min} = 0.05~{\rm mJy~kpc^2}$. Note that this value of $L_{\rm 1400, min}$ is higher than the presently known lowest value of $L_{1400} = 0.01~{\rm mJy~kpc^2}$ for the disk pulsar J1741-2054 (Camilo \textit{et~al.}\cite{crr09}). Bagchi \textit{et~al.} did not find any correlation between the predicted number of pulsars in different GCs and GC parameters, but with the improved estimates of stellar encounter rates, Bahramian \textit{et~al.}\cite{bhsg13} demonstrated that the number of pulsars as predicted by Bagchi \textit{et~al.} was correlated with the stellar encounter rates.

Chennamangalam \textit{et~al.}\cite{jayanth13} employed a Bayesian technique to further constrain the parameters of the lognormal distribution. They treated pulsar populations in different globular clusters separately unlike most of the previous works where people studied the total pulsar (or recycled pulsar) population in all clusters. Moreover, they first performed their study in the flux domain and then translated the results to the luminosity domain. In addition to flux values of individual pulsars, they used the values of total diffused flux also. They studied three clusters, Terzan 5, 47 Tuc and M 28, as these are the top three according to the number of pulsars with known flux estimates, and total diffuse flux values are also known. Note that although they did not explicitly put any condition on the spin period of the pulsars, their dataset contained only recycled (for a few cases though mildly) pulsars. Table \ref{tb:jayanth13} summarizes the data and priors they used and the posteriors they obtained. The values of flux of individual pulsars and total diffuse flux values ($S_{\rm 1400, tot}$) for each GC were taken from the literature. See Table 1 of Bagchi \textit{et~al.} for individual flux values and spectral indices. The priors in the distance ($d$) were taken as Gaussians centered at best distance estimates available, and standard deviations as the errors reported. The priors for the total number of pulsars were uniform distributions between the upper and lower limits where the lower limits were the total number of pulsars with measured flux values and the upper limits were obtained from the upper limits obtained by Bagchi \textit{et~al.} for the lognormal distribution with $\mu = -1.1$ and $\sigma = 0.9$, plus additional 150 percent of those values. The priors in $\mu$ and $\sigma$ were also uniform distributions in specified ranges - the wide ranges were taken from Bagchi \textit{et~al.} and the narrow ranges were taken from Ridley and Lorimer. The choices of the narrow ranges were based on the conclusion of Bagchi \textit{et~al.} that the luminosity distribution for the disk pulsars and the cluster pulsars are the same (Ridley and Lorimer studied disk pulsars). For each cluster, the prior in $S_{min}$ was taken as uniform in the range $0$ to the minimum of the measured pulsar fluxes for that cluster. Interestingly, osteriors in $\mu$ and $\sigma$ were different for different clusters. Although Chennamangalam \textit{et~al.} succeeded to narrow down the allowed ranges of $\mu$ and $\sigma$, their allowed ranges for number of pulsars were much larger than those obtained by Bagchi \textit{et~al}. There is scope for improving both these analyses by including more data points (whenever more flux values will be available) and by incorporating uncertainties in flux measurements.

\begin{table*}[h]
\tbl{Data (other than individual flux values), priors, and posteriors in Chennamangalam \textit{et~al.}\cite{jayanth13}. References are a: Fruchter and Goss\cite{fg00} , b:  Ortolani \textit{et~al.}\cite{obb07} , c: McConnell and Deshpande\cite{mdca04} , d: Woodley, Goldsbury, \& Kalirai\cite{wgk12} , e: Kulkarni \textit{et~al.}\cite{kgw90} , f: Servillat \textit{et~al.}\cite{shh12} .}
{\begin{tabular}{ |l|c|c|c|c|c|c|c|c|}
\hline
 & Data &\multicolumn{4}{ |c| }{Priors}  & \multicolumn{3}{ |c| }{Posteriors}\\
\cline{2-9} 
 & $S_{\rm 1400, tot}$ & $d$  & $N$ & $\mu$ & $\sigma$ & N & $\mu$ & $\sigma$\\ 
   &  & (Gaussian) & (uniform)  & (uniform) & (uniform) &  &  & \\ 
GC  & (kpc) & (mJy) &  &  &  &  &  & \\ 
\hline
\multirow{2}{*}{Terzan 5} & \multirow{2}{*}{$ 5.2 ^{a}$} & \multirow{2}{*}{$5.5 \pm 0.9 ^{b}$}  & \multirow{2}{*}{[25, 500]} & [-2.0, 0.5] & [0.2, 1.4] & $142^{+310}_{-110}$ & $-1.2^{+1.4}_{-0.8}$ & $1.0^{+0.3}_{-0.4}$\\
\cline{5-9}
  &   &  &  & [-1.19, -1.04] & [0.91, 0.98] & $147^{+112}_{-65}$ & $-1.12^{+0.08}_{-0.07}$ & $0.94^{+0.03}_{-0.03}$\\ \hline
\multirow{2}{*}{47 Tuc} & \multirow{2}{*}{$ 2.0 ^{c}$} & \multirow{2}{*}{$4.69 \pm 0.17^{d}$}  & \multirow{2}{*}{[14, 225]} & [-2.0, 0.5] & [0.2, 1.4] & $39^{169}_{-25}$ & $-0.6^{+0.9}_{-1.3}$ & $0.7^{+0.4}_{-0.4}$\\
\cline{5-9}
  &    &   & & [-1.19, -1.04] & [0.91, 0.98] & $83^{+54}_{-35}$ & $-1.13^{+0.08}_{-0.07}$ & $0.94^{+0.04}_{-0.03}$\\ \hline
\multirow{2}{*}{M 28} & \multirow{2}{*}{$ 1.8 ^{e}$} & \multirow{2}{*}{$5.5 \pm 0.3^{f}$}  & \multirow{2}{*}{[9, 400]} & [-2.0, 0.5] & [0.2, 1.4] & $198^{+191}_{-169}$ & $-1.3^{+1.1}_{-0.7}$ & $0.8^{+0.3}_{-0.3}$\\
\cline{5-9}
  &   &  &  & [-1.19, -1.04]  & [0.91, 0.98] & $100^{+91}_{-52}$ & $-1.13^{+0.09}_{-0.06}$ & $0.94^{+0.04}_{-0.03}$\\
\hline
\end{tabular}
\label{tb:jayanth13}}
\end{table*}


\section{Summary and Conclusion}

In this review, I tried to cover all the significant works on luminosities of radio pulsars. The main open question at this point is that if there is no difference between the luminosity distribution for normal isolated pulsars in the Galactic disk and recycled pulsars in GCs (as found by Bagchi \textit{et~al.}), then why the difference in the luminosity distributions of pulsars in different GCs has been observed by Chennamangalam \textit{et~al.}? Insufficient data may have biased any one or both of these studies. The second question is that whether luminosities of radio pulsars depend on the values of $P_{s}$ and $\dot{P}_s$, and if yes then what is the exact form and what is the physical explanation? Why the lognormal function provides a better fit than the conventional power-law? What are the exact parameters for the lognormal distribution? We do not know answers to these questions and do not know how much the past studies have been affected by observation limitations. So continuous study on this topic is needed.


\section*{Acknowledgments}

The author was supported by a Research Challenge Grant to the WVU Center for Astrophysics by the West Virginia EPSCoR
foundation, and thanks Duncan Lorimer for introducing her to this field and Jayanth Chennamangalam for useful comments on the manuscript.


\end{document}